\documentclass{article}
\usepackage{graphicx}  
\usepackage{amsmath}   
\usepackage[compress]{cite}
\usepackage{amssymb}   
\usepackage{bm} 
\usepackage{dcolumn}
\usepackage{color}
\usepackage{mathrsfs}
\usepackage{amsfonts}
\usepackage{varioref}
\RequirePackage[colorlinks,citecolor=blue,urlcolor=magenta,linkcolor=blue]{hyperref}
\labelformat{section}{Section #1} 
\labelformat{subsection}{Section #1} 
\labelformat{subsubsection}{Section #1}
\labelformat{subsubsubsection}{Section #1}
\labelformat{equation}{Eq.~(#1)} 
\labelformat{figure}{Fig.~#1} 
\labelformat{subfigure}{Fig.~\thefigure#1} 
\labelformat{table}{Tab.~#1} 
\labelformat{appendix}{Appendix #1}

\def\EH{Einstein-Hilbert }
\def\LL{Lanczos-Lovelock }
\def\gr{general relativity}
\def\gb{Gauss-Bonnet }
\title{Spherically symmetric brane in a bulk of f(R) and Gauss-Bonnet Gravity}
\author{Sumanta Chakraborty 
\footnote{sumantac.physics@gmail.com;~~sumanta@iucaa.in}\\
{\small {IUCAA, Post Bag 4, Ganeshkhind, Pune University Campus, Pune 411 007, India}}\\
and \\
Soumitra SenGupta
\footnote{tpssg@iacs.res.in}\\
{\small{Department of Theoretical Physics}}\\
{\small {Indian Association for the Cultivation of Science, Kolkata-700032, India}}}
\begin{document}

\maketitle
\begin{abstract}
Effective gravitational field equations on a four dimensional brane embedded in a five dimensional bulk have been considered. Using the Einstein-Hilbert action along with the Gauss-Bonnet correction term, we have derived static spherically symmetric vacuum solution to the effective field equations, first order in the Gauss-Bonnet coupling parameter. The solution so obtained, has one part corresponding to general relativity with an additional correction term, proportional to the Gauss-Bonnet coupling parameter. The correction term modifies the spacetime structure, in particular, the location of the event horizon. Proceeding further, we have derived effective field equations for $f(R)$ gravity with Gauss-Bonnet correction term and a static spherically symmetric solution has been obtained. In this case the Gauss-Bonnet term modifies both the event and cosmological horizon of the spacetime. There exist another way of obtaining the brane metric --- expanding the bulk gravitational field equations in the ratio of 
bulk to brane curvature scale and assuming a separable bulk metric ansatz. It turns out that static, spherically symmetric solutions obtained from this perturbative method can be matched exactly, with the solutions derived earlier. This will hold for Einstein-Hilbert plus Gauss-Bonnet as well as for f(R) with the Gauss-Bonnet correction. Implications of these results are discussed.
\end{abstract}
\section{Introduction}\label{GB_Intro}

Higher dimensional spacetime is a natural working stage for string theory \cite{Polchinski:1998rq,Polchinski:1998rr,Horava:1995qa}. One of the string inspired models implementing higher dimensional spacetime manifold in order to address a possible resolution of the long standing hierarchy problem is the brane world model \cite{ArkaniHamed:1998rs,Antoniadis:1998ig}. Initial works in this direction assumed the extra spatial dimensions to be flat and compact --- the volume of the extra dimensions suppress the Planck scale in the bulk (the full higher dimensional spacetime) to TeV scale on the brane (which is the 4-dimensional hypersurface we live in). The problem with this approach is that it does not include gravity, which being universal (all particles feels it) and long range must propagate on the bulk (for another novel justification see \cite{Dadhich:2015lra}). This problem was remedied immediately after, in \cite{Randall:1999ee} by invoking gravity in the bulk and thereby 
warping the extra dimensions, such that the warp factor reduces the Planck scale to TeV scale on the visible brane. Since then there have been numerous investigations in the context of brane world scenario, e.g., in cosmology \cite{Chakraborty:2013ipa,Lukas:1998qs,Binetruy:1999ut,Csaki:1999jh,Csaki:1999mp,Brax:2004xh,Ida:1999ui,Guha:2009pba,
Chakraborty:2007ad,Nojiri:2002hz,Cembranos:2008kg}, in particle phenomenology \cite{Chakraborty:2014zya,Davoudiasl:2000wi,Davoudiasl:1999tf,Djouadi:2005gi}, black hole formation and characterization \cite{Chamblin:1999by,Garriga:1999yh,Chakraborty:2007zh} and dark matter phenomenology \cite{Cembranos:2003mr,Cembranos:2011cm,Cembranos:2016jun,Chakraborty:2015zxc}.

The existence of extra dimensions leads to effective gravitational field equations on the brane which inherits additional contributions from the bulk. The effective gravitational field equations on the brane due to Einstein bulk was first derived in \cite{Shiromizu:1999wj} using Gauss-Codazzi equation and static, spherically symmetric solutions were derived subsequently in \cite{Dadhich:2000am,Harko:2004ui}. The above formalism has recently been generalized for arbitrary dimensions in the context of \gr\ in \cite{Chakraborty:2015bja}. Apart from this method of obtaining effective field equations by projecting bulk equation on the brane, there is another perturbative way. In this perturbative method, the ratio of bulk to brane curvature is treated as the perturbation parameter. In this method one solves for bulk equations using a factorizable metric ansatz and in an iterative manner. The bulk field equations for $n$th order in the perturbation theory is solved, using solution at $(n-1)$th order. Subsequently, 
the solution for $n$th order is used to obtain solution for $(n+1)$th order \cite{Shiromizu:2002qr,Kanno:2002iaa,Kanno:2002kz,Chakraborty:2015xja}.

In all the above attempts it has been assumed that bulk spacetime is well described by \gr. However due to Planck scale physics on the bulk, it is natural to treat \gr\ as only a zeroth order approximation and thus it should inherit higher order corrections. Two most natural candidates in this direction are, \LL gravity and $f(\mathcal{R})$ gravity. \LL gravity is unique in the sense that even though it has higher order curvature corrections, the field equations are only second order in the dynamical variables \cite{Padmanabhan:2013xyr}. Moreover from the thermodynamic perspectives as well \LL gravity is preferred \cite{Chakraborty:2014rga,Chakraborty:2014joa,Chakraborty:2015wma,Chakraborty:2015hna,Chakraborty:2015kva}. The first correction due to higher order \LL theories of gravity to the \EH Lagrangian corresponds to the \gb term. The behavior of effective field equations in presence of the \gb term in the bulk, has been studied in \cite{Maeda:2003vq} \footnote{To avoid confusion we should mention that 
solution to the bulk Einstein-Gauss-Bonnet field equations has been obtained much earlier in \cite{Boulware:1985wk,Wiltshire:1985us,Wiltshire:1988uq}. In this work we are interested in spherically symmetric vacuum brane solutions, which to our knowledge has not been derived earlier.}. On the other hand the perturbative scheme of obtaining the gravitational field equations on the brane for \gb gravity in the bulk has been detailed in \cite{Kobayashi:2006jw}. In this work we will use both these formalisms in order to obtain static, spherically symmetric vacuum solutions \emph{on the brane} with \gb term in the bulk. We will also discuss validity of these approaches as well as their similarity and differences.

On the other hand, the $f(\mathcal{R})$ theories of gravity are important from cosmological perspective. Both the late time cosmic acceleration as well as the inflationary scenario are very well described by invoking a $f(\mathcal{R})$ term in the gravitational action. However in order to become a viable model the $f(\mathcal{R})$ theories should pass the local gravity tests --- perihelion precession of Mercury and bending of light. The effect of local gravity tests are imprinted on the Eddington parameters routinely used in Post Newtonian calculations and hence one should compute these Eddington parameters using Post Newtonian formulations keeping in mind that essentially $f(\mathcal{R})$ gravity is a Brans-Dicke theory with $\omega =0$. There have been recent claims that for $f(\mathcal{R})$ gravity, the Eddington parameter $\gamma$ has a value too far from the observational result, which is $\gamma =1+(2.1\pm 2.3)\times 10^{-5}$ \cite{Olmo:2005hc,Olmo:2005zr,Allemandi:2005tg,Bertotti:2003rm}. Shortly 
after it was realized, that the above result is not a generic one, for some non-trivial choices of $f(\mathcal{R})$ (which is equivalent to some non-trivial potential in scalar-tensor representation) one can still have $\gamma$ within experimental bounds. This suggests that Solar System experiments do not exclude the viability of higher order gravity theories even at scales shorter than the cosmological ones. This constraint on $\gamma$, on performing a post Newtonian calculation gets transferred to the quantity $f''(\mathcal{R})^{2}$, which in turn leads to constraints on the parameters of the model. Combining both perihelion shift of Mercury and Very Long Baseline Interferometry one can show that models like $f(R)=f_{0}R^{n}$, $R+aR^{2}$, $R+\mu /R$, $A\log R$ are all consistent with local gravity tests provided the parameters $f_{0},a,\mu ,A$ satisfy certain bounds. The weak field limit indicates that several higher order Lagrangians are viable on the Solar System scales even though the Solar System 
experiments pose rather tight constraints on the values of coupling constants. Hence it seems reasonable to affirm that extended gravity theories cannot be ruled out, definitively, 
using Solar System experiments \cite{Capozziello:2005bu,Capozziello:2007ms,Sotiriou:2005xe,Capozziello:2006jj}. 
Given the fact that $f(\mathcal{R})$ theories of gravity with certain restriction 
on its parameters pass through both cosmological and local gravity tests, 
can be a strong candidate for explaining gravity at high energies 
\cite{Nojiri:2010wj,Sotiriou:2008rp,DeFelice:2010aj,Ayuso:2014jda,delaCruz-Dombriz:2013gfa,Abebe:2013zua,Resco:2016upv,
Cembranos:2011sr,delaCruz-Dombriz:2015tye,delaCruzDombriz:2009et,delaCruzDombriz:2012xy}.

Even though derivation of gravitational field equations on the brane via Gauss-Codazzi formalism or the perturbative expansion in bulk to brane curvature ratio is well known for Einstein-Gauss-Bonnet gravity, it has not been derived earlier for $f(\mathcal{R})$ gravity theories with \gb correction term. In this work we have filled this gap. Starting from a bulk action which depends on both $f(\mathcal{R})$ gravity and the \gb term, we have derived the brane gravitational field equations by both the procedures sketched above.

Unlike the situation with \gb term, where one can obtain unique black hole solutions, with $f(\mathcal{R})$ gravity theories obtaining new black hole solutions are difficult, modulo no-hair theorems. However the no-hair theorems depend crucially on the fall-off behavior of the scalar field, i.e., the scalar field must vanish at infinity. When this is not the case one can have non-trivial solutions even in the context of $f(\mathcal{R})$ gravity (or, equivalently scalar-tensor theories) as well \cite{Capozziello:2007wc,Multamaki:2006zb,Multamaki:2006ym,Bhattacharya:2015oma,Paliathanasis:2011jq}. We will consider precisely such a situation in which the scalar field does not vanish at infinity vis-\'{a}-vis in the context of brane world and shall obtain spherically symmetric solutions to the effective field equations derived in this work.

The paper is organized as follows: We start with a brief introduction to the effective field equations on the brane with Einstein-Gauss-Bonnet gravity in the bulk and derive the respective ones for bulk $f(\mathcal{R})$ gravity with \gb correction term in \ref{GB_Review}. Then we have obtained spherically symmetric solutions to the effective field equations, derived for \gb correction in \ref{GB_SphBrane} and for $f(\mathcal{R})$ gravity with \gb correction in \ref{frgb}. Subsequently in \ref{pergb} we have described the perturbative formalism and the equivalence with results obtained in earlier sections. Finally, we conclude with a discussion on our results.

We have set the fundamental constants $c$ and $\hbar$ to unity and shall work with mostly positive signature of the metric, i.e., $(-,+,+,+,\ldots)$. The Latin indices $a,b,c,\ldots$ run over the bulk spacetime, while the Greek indices $\mu,\nu,\ldots$ spans the brane spacetime.
\section{Effective Gravitational Field Equations: A Brief Review}\label{GB_Review}

Given a gravity theory in the bulk spacetime, our main aim is to obtain and solve the effective field equations on the brane. Before plunging into solving the field equations, we provide a brief introduction to the basic formalism of getting the equations for the case of \EH action with \gb correction term. Later we have also \emph{derived} the gravitational field equations on the brane with both $f(\mathcal{R})$ and \gb gravity in the bulk. 
\paragraph*{Einstein-Gauss-Bonnet Field Equations on Brane: Gauss-Codazzi Formalism} The bulk spacetime is taken to be $5$-dimensional with both \EH and \gb terms such that the action is described by,
\begin{align}\label{GB_01}
\mathcal{A}_{\rm{bulk}}=\int _{\mathcal{M}}d^{5}x\sqrt{-g}\left[\frac{1}{2\kappa _{5}^{2}}\left(\mathcal{R}+\alpha \mathcal{L}_{\rm{GB}}\right)+\mathcal{L}_{\rm{m}}\right]
\end{align}
where $\kappa _{5}^{2}$ represents the $5$-dimensional gravitational constant, $\mathcal{L}_{\rm{m}}$ stands for the matter Lagrangian and $\mathcal{L}_{\rm{GB}}$ represents the \gb correction term and has the expression,
\begin{align}\label{GB_02}
\mathcal{L}_{\rm{GB}}=\mathcal{R}^{2}-4\mathcal{R}_{ab}\mathcal{R}^{ab}+\mathcal{R}_{abcd}\mathcal{R}^{abcd}
\end{align}
In the above expressions $\mathcal{R}$, $\mathcal{R}_{ab}$ and $\mathcal{R}_{abcd}$ are the bulk curvature tensor and its contractions. $\alpha$ appearing in the above expressions is known as the \gb coupling constant. Varying the brane plus bulk action the gravitational field equations on the bulk turns out to be \cite{gravitation},
\begin{align}\label{GB_03}
\mathcal{G}_{ab}+\alpha \mathcal{H}_{ab}=\kappa _{5}^{2}\left[\mathcal{T}_{ab}+\tau _{ab}\delta(\Sigma) \right]
\end{align}
where we have used the following definitions,
\begin{align}
\mathcal{G}_{ab}&=\mathcal{R}_{ab}-\frac{1}{2}g_{ab}\mathcal{R}
\label{GB_04a}
\\
\mathcal{H}_{ab}&=2\left(\mathcal{R}\mathcal{R}_{ab}-2\mathcal{R}_{ac}\mathcal{R}^{c}_{b}-2\mathcal{R}^{cd}\mathcal{R}_{acbd}+\mathcal{R}_{a}^{~cde}\mathcal{R}_{bcde}\right)-\frac{1}{2}g_{ab}\mathcal{L}_{\rm{GB}}
\label{GB_04b}
\end{align}
and $\mathcal{T}_{ab}$ is the matter energy-momentum tensor originating from the bulk matter action $\mathcal{L}_{\rm{m}}$ with $\tau _{ab}$ being brane energy momentum tensor (the brane is located at $\Sigma =0$, which explains the Dirac delta function associated with $\tau _{ab}$). 

Using the Gauss-Codazzi equation \cite{gravitation}, connecting bulk curvature tensors to the brane curvature tensors, the effective equations on the brane can be obtained as \cite{Maeda:2003vq},
\begin{align}\label{GB_05}
G_{\mu \nu}&+E_{\mu \nu}-KK_{\mu \nu}+K_{\mu \rho}K^{\rho}_{\nu}+\frac{1}{2}\left(K^{2}-K_{\alpha \beta}K^{\alpha \beta}\right)h_{\mu \nu}+\alpha \left(H^{(1)}_{\mu\nu}+H^{(2)}_{\mu\nu}
+H^{(3)}_{\mu\nu}\right)
\nonumber
\\
&=\frac{2\kappa _{5}^{2}}{3}\left[\left\lbrace \mathcal{T}_{ab}e_{\mu}^{a} e_{\nu}^{b}+\left(\mathcal{T}_{ab}n^{a}n^{b}-\frac{1}{4}\mathcal{T}\right)h_{\mu\nu}\right\rbrace +\frac{\alpha}{3+\alpha M}\left(M_{\mu\nu}-\frac{1}{4}Mh_{\mu\nu}\right)\mathcal{T}_{ab}h^{ab}\right]
\end{align}
In the above expression, $E_{\mu \nu}$ stands for the electric part of the bulk Weyl tensor projected on the brane, $K_{\mu \nu}$ stands for the extrinsic curvature of the brane with $K$ representing its trace. The three additional tensors $H^{(1)}_{\mu \nu}$, $H^{(2)}_{\mu \nu}$ and $H^{(3)}_{\mu \nu}$ appearing in \ref{GB_05} stands for the various quadratic combinations of the curvature tensor, Weyl tensor and derivatives of extrinsic curvature. Since the expressions for these tensors are quiet involved, we have postponed their expressions till \ref{GBAPP_01} (for detailed expressions see \ref{App_E02}, \ref{App_E03} and \ref{App_E04} in \ref{GBApp_0101}). Note that if we substitute $\alpha =0$, then we would retrieve the effective equations for the \EH action.

As a final ingredient to the derivation of effective field equations as presented in \ref{GB_05} we consider the Israel Junction conditions on the brane. For \gr\ the junction conditions are simple, the discontinuity in the extrinsic curvature should be related to brane energy-momentum tensor. However when applied to \gb gravity the junction condition becomes complicated, along with the extrinsic curvature it will depend on curvatures as well. It essentially follows from the boundary term for \gb action first derived in \cite{Myers:1987yn} and then applied to braneworld models in \cite{Gravanis:2002wy,Davis:2002gn,Maeda:2003vq}, which reads
\begin{align}
2\langle K_{\mu \nu}-Kh_{\mu \nu}\rangle +4\alpha \langle 3J_{\mu \nu}-Jh_{\mu \nu}+2K^{\alpha \beta}P_{\mu \alpha \nu \beta}\rangle=-\kappa _{5}^{2} \tau _{\mu \nu}
\end{align}
where $P_{\mu \alpha \nu \beta}$ is the projection of $P_{abcd}$, divergence free part of Riemann tensor on the brane and has the following expression
\begin{align}
P_{abcd}=R_{abcd}+\left(R_{bc}g_{da}-R_{bd}g_{ca}\right)-\left(R_{ac}g_{db}-R_{ad}g_{bc}\right)+R\left(g_{ac}g_{bd}-g_{ad}g_{bc}\right)
\end{align}
$\tau _{ab}$ is the brane energy momentum tensor defined in \ref{GB_03}. The tensor $J_{\mu \nu}$ is constructed out of extrinsic curvature of the brane hypersurface only and can be written as,
\begin{align}
J_{\mu \nu}=\frac{1}{3}\left(2KK_{\mu \rho}K^{\rho}_{\nu}+K_{\rho \sigma}K^{\rho \sigma}K_{\mu \nu}-2K_{\mu \rho}K^{\rho \sigma}K_{\sigma \nu}-K^{2}K_{\mu \nu}\right)
\end{align}
with $J$ representing its trace and $\langle X\rangle$ stands for $X$ evaluated on either side of the brane. These junction conditions have to be used in parallel with the effective field equations presented in \ref{GB_05} in order to obtain complete dynamics. 
\paragraph*{$\boldsymbol{f(\mathcal{R})}$-Gauss-Bonnet Field Equations on Brane: Gauss-Codazzi Formalism} The above effective field equations on a brane have been derived on the premise when the bulk spacetime is endowed with \EH Lagrangian plus a \gb correction term. Given the current significance of the $f(\mathcal{R})$ gravity model another important situation arises if one considers a $f(\mathcal{R})$ term, where $f$ is an arbitrary function of $\mathcal{R}$, the Ricci scalar of the bulk spacetime along with the \gb correction term. The relevant action becomes,
\begin{align}\label{GB_06}
\mathcal{A}_{\rm{bulk,f}}=\int _{\mathcal{M}}d^{5}x\sqrt{-g}\left[\frac{1}{2\kappa _{5}^{2}}\left\lbrace f\left(\mathcal{R}\right)+\alpha \mathcal{L}_{\rm{GB}}\right\rbrace +\mathcal{L}_{\rm{m}}\right]
\end{align}
Following the same procedure, using the Gauss-Codazzi equations one can relate brane curvature tensors to those in the bulk and finally project the bulk gravitational field equations on the brane. After all these steps the effective gravitational field equations on the brane finally turn out to be (see \cite{Maeda:2003vq} and \cite{Chakraborty:2014xla})
\begin{align}\label{GB_07}
G_{\mu \nu}&+E_{\mu \nu}-F(\mathcal{R})h_{\mu \nu}-KK_{\mu \nu}+K_{\mu \rho}K^{\rho}_{\nu}+\frac{1}{2}\left(K^{2}-K_{\alpha \beta}K^{\alpha \beta}\right)h_{\mu \nu}+\alpha \left(\hat{H}^{(1)}_{\mu\nu}+\hat{H}^{(2)}_{\mu\nu}
+\hat{H}^{(3)}_{\mu\nu}\right)
\nonumber
\\
&=\frac{2\kappa _{5}^{2}}{3}\left[\left\lbrace \mathcal{T}_{ab}e_{\mu}^{a} e_{\nu}^{b}+\left(\mathcal{T}_{ab}n^{a}n^{b}-\frac{1}{4}\mathcal{T}\right)h_{\mu\nu}\right\rbrace +\frac{\alpha}{3+\alpha M}\left(M_{\mu\nu}-\frac{1}{4}Mh_{\mu\nu}\right)\mathcal{T}_{ab}h^{ab}\right]
\end{align}
Let us explain various terms in the above expression. $G_{\mu \nu}$ stands for the standard Einstein tensor and $\mu, \nu$ refers to brane indices. $E_{\alpha \beta}$ represents the electric part of bulk Weyl tensor $C_{abcd}$ projected on the brane. The tensors $H^{(1)}_{\mu \nu}$, $H^{(2)}_{\mu \nu}$ and $H^{(3)}_{\mu \nu}$ contain various quadratic combinations of the curvature, Weyl tensor and derivatives of extrinsic curvature tensor. Since the expressions for these tensors are quiet involved, we have presented them in \ref{App_E02}, \ref{App_E03} and \ref{App_E04} of \ref{GBApp_0101} respectively. Furthermore, the term $M_{\mu \nu}$ corresponds to $R_{\mu \nu}-KK_{\mu \nu}+K_{\mu \alpha}K^{\alpha}_{\nu}$ and $M$ is just the trace of $M_{\alpha \beta}$, i.e., $M=h^{\alpha \beta}M_{\alpha \beta}$. Also the term $F(\mathcal{R})$ appearing on the left hand side of \ref{GB_07} is connected to the Lagrangian term $f(\mathcal{R})$ through the following relation,
\begin{align}\label{GB_08}
F(\mathcal{R})=\left[\frac{1}{4}\frac{f(\mathcal{R})}{f'(\mathcal{R})}-
\frac{1}{4}\mathcal{R}-\frac{2}{3}\frac{\square f'(\mathcal{R})}{f'(\mathcal{R})}+\frac{2}{3}\frac{\nabla _{a}\nabla _{b}f'(\mathcal{R})}
{f'(\mathcal{R})}n^{a}n^{b}\right]_{\Sigma}
\end{align}
Here $\Sigma$ in the subscript of \ref{GB_08} signifies that the quantity is evaluated on the hypersurface $\Sigma$, i.e., on the brane. Note that for $f(\mathcal{R})=\mathcal{R}$, $F(\mathcal{R})$ vanishes identically and we get back \ref{GB_05}. 

Just like the case of \gb gravity in the case of $f(\mathcal{R})$ gravity as well junction conditions are of quiet importance. Recently, there have been lots of work in this direction and most of them have discussed the junction conditions to some extent. We would not repeat the full analysis but shall provide a brief argument. The field equations for $f(\mathcal{R})$ gravity can be written as equivalent to Einstein's equations with the additional curvature contributions as an energy momentum tensor. Thus the junction condition becomes conditions on extrinsic curvature only, with the right hand side containing projection of the additional bulk energy momentum tensor on the brane, see for example \cite{Borzou:2009gn,Haghani:2013oma,Carames:2012gr,Haghani:2012zq,Apostolopoulos:2010jg}. 
\paragraph*{Einstein-Gauss-Bonnet Field Equations on Brane: Perturbative Method} The other approach to get field equations on the brane requires the bulk equations to be expanded perturbatively in the ratio of bulk to brane curvature. Perturbative expansion for the action presented in \ref{GB_01} leads to the following field equations on the brane \cite{Kobayashi:2006jw}
\begin{align}\label{GB_09}
G_\mu^{\nu}&=\frac{\kappa_{5}^{2}}{\ell(1+\beta)}T_{\mu}^{~\nu}+\frac{2}{\ell}\frac{1-\beta}{1+\beta}{}^{(2)}\bar\chi_\mu^{~\nu}
+\frac{(1-3\beta)\ell^2}{1+\beta}{\cal P}_\mu^{~\nu}+\frac{\beta\ell^2}{1+\beta}C_{\mu\alpha}^{~~~\nu\beta}R_{\beta}^{~\alpha}-\frac{\beta\ell^2}{3}\left[{\cal W}_\mu^{~\nu}-\frac{7}{16}\delta_\mu^{~\nu} {\cal  W} \right].
\end{align}
where $\beta =(4\alpha /\ell ^{2})$ is a dimensionless quantity, with $\alpha$ being the \gb coupling constant and $\ell$ being the bulk curvature radius. The tensor $T_{\mu \nu}$ on the right hand side of low energy effective field equations is the energy-momentum tensor on the brane, $R_{\alpha \beta}$ is the Ricci tensor on the brane and $C_{\alpha \beta \mu \nu}$ is the four-dimensional Weyl tensor. The other tensors have a much involved expressions and have been defined explicitly in \ref{GBApp_0101} (in particular \ref{Low01} to \ref{GB_New_04}).
\paragraph*{$\boldsymbol{f(\mathcal{R})}$-Gauss-Bonnet Field Equations on Brane: Perturbative Method} Finally the above result, being derived for the \EH action plus the \gb correction term, can be generalized in a straightforward manner to $f(\mathcal{R})$ plus the \gb correction term. In this case the gravitational field equations on the brane will take the following form,
\begin{align}\label{GB_New05}
G_\mu^{\nu}-F(\mathcal{R})\delta ^{\nu}_{\mu}&=\frac{\kappa_{5}^{2}}{\ell(1+\beta)}T_{\mu}^{~\nu}+\frac{2}{\ell}\frac{1-\beta}{1+\beta}{}^{(2)}\bar\chi_\mu^{~\nu}
+\frac{(1-3\beta)\ell^2}{1+\beta}{\cal P}_\mu^{~\nu}
\nonumber
\\
&+\frac{\beta\ell^2}{1+\beta}C_{\mu\alpha}^{~~~\nu\beta}R_{\beta}^{~\alpha}-\frac{\beta\ell^2}{3}\left[{\cal W}_\mu^{~\nu}-\frac{7}{16}\delta_\mu^{~\nu} {\cal  W} \right].
\end{align}
Here also $\beta$ is a dimensionless parameter, $(4\alpha /\ell ^{2})$, where $\alpha$ is the \gb coupling constant and $\ell$ being the bulk curvature radius. The tensor $T_{\mu \nu}$ on the right hand side of the field equations is the energy-momentum tensor, $R_{\alpha \beta}$ is the Ricci tensor  and $C_{\alpha \beta \mu \nu}$ is the Weyl tensor on the brane. Further the quantity $F(\mathcal{R})$ has the expression as detailed in \ref{GB_08}. Among the others $\mathcal{W}^{\mu}_{\nu}$ is quadratic in the Weyl tensor, $^{(2)}\bar{\chi}_{\mu \nu}$ depends on the quadratic terms of brane curvature and Weyl tensors, as well as on tensors originating from integration of bulk equations over the extra dimension. The detailed expressions has been provided in \ref{GBApp_0101} from \ref{Low01} to \ref{GB_New_04}.
\paragraph*{The two approaches: A Comparative Study} Let us briefly discuss the pros and cons of the two methods discussed above. The first method of projecting the bulk equations using Gauss-Codazzi relations is exact. While the second one is a perturbative method --- the bulk equations are expanded in terms of bulk to brane curvature ratio. Hence the second method relies heavily on the fact that $\textrm{bulk~curvature~scale}\ll\textrm{brane~curvature~scale}$, which is satisfied at the mesoscopic (or intermediate) energy scales we are interested in. The validity of the perturbative method is manifested in our analysis, viz., the solution obtained through the first technique \emph{exactly} matches with that obtained by the second, both being first order in the \gb coupling parameter. The equivalence at the energy scales of interest not only holds for Einstein-Gauss-Bonnet but for $f(\mathcal{R})$-Gauss-Bonnet gravity as well, which we have illustrated in later sections. Thus one might use 
the 
perturbative technique, which is simpler compared to the other exact formulation as long as the bulk curvature radius is much smaller compared to that on the brane.

As a final remark, we will consider the matter content of our theory. We have bulk energy momentum tensor $\mathcal{T}_{ab}$ and its projection on the brane along with the brane energy momentum tensor $\tau _{ab}$. We will assume that there is no matter field on the bulk, except for a negative bulk cosmological constant $\Lambda$, while the energy momentum tensor on the brane is constructed out of brane tension $\Sigma$ alone. Thus we have a vacuum brane. To linear order in \gb coupling parameter $\alpha$, what we will be interested in, the effect of these two will be a combination $-\vert \Lambda \vert +(1/6)\kappa _{5}^{2}\Sigma ^{2}$ \cite{Shiromizu:1999wj,Maeda:2003vq,Torii:1996yi}. Choosing the brane tension appropriately this term can be made to vanish, which is equivalent to assuming existence of a small cosmological constant. We will work under these assumptions for simplicity, but the results can be generalized straightforwardly to include effective brane cosmological constant as well. We will now 
try to solve the above equations and obtain vacuum spherically symmetric solutions.
\section{Static, Spherically Symmetric Vacuum Brane with Einstein-Gauss-Bonnet Gravity in the Bulk}\label{GB_SphBrane}

In this section we will be concentrating on \EH action with a \gb correction term. As mentioned in the introduction itself, we are interested in vacuum solutions of the effective field equations, which for this particular situation are given in \ref{GB_05}. Even though the effective equations in \ref{GB_05} looks rather involved, with additional symmetries, e.g., staticity, spherical symmetry reduces the effective equations to a tractable form. Further the spacetime is taken to be vacuum, it immediately follows that the brane energy-momentum tensor identically vanishes. This in turn implies vanishing of the extrinsic curvature, due to the Junction conditions. For this special class of static and spherically symmetric spacetime the line element can be written as,
\begin{align}\label{GB_10}
ds^{2}=-e^{\nu (r)}dt^{2}+e^{\lambda (r)}dr^{2}+r^{2}d\Omega ^{2}
\end{align}
At this stage it would be interesting to briefly mention existing results in the literature. Gauss-Bonnet gravity in the bulk was an active area research and quiet a lot of results have been derived. All these works can be divided broadly into two classes:
\begin{itemize}
\item people have started from the famous Deser-Boulware solution in \gb bulk, where the bulk metric is spherically symmetric. Subsequently, the brane was assumed to have a time dependent position in the bulk such that the metric reduces to cosmological Friedmann ansatz on the brane. Then by taking the energy momentum tensor on the brane to be perfect fluid, using junction conditions one can immediately derive the Hubble parameter. This closes the brane-bulk system, with spherically symmetric bulk and cosmological brane \cite{Davis:2002gn,Deruelle:2003ur,Deruelle:2000ge,Gravanis:2002wy,Cho:2001su,Meissner:2000dy} (Note that one should carefully regularize the delta function present in the junction condition, otherwise one might led to inconsistencies \cite{Germani:2002pt}).
\item The same result can also be obtained but from a different perspective. One can start with a metric ansatz such that the brane is located at a fixed point and have cosmological solution. Then using integrability of the field equations one arrives at various classes of solutions. Among them one class leads to Deser-Boulware solution in the bulk and cosmological solution in the brane with identical Hubble parameter as in the above situation. However they could obtain other non-trivial solutions as well \cite{Charmousis:2002rc,Abdesselam:2001ff,Mavromatos:2000az,Torii:1996yi,Lidsey:2002zw}.
\end{itemize}
This should be contrasted from our aim in this work. Both the approaches lead to cosmological solutions on the brane \emph{not} spherically symmetric black hole solution. Secondly, use of Junction condition as one of the field equations is possible only in cosmology which is dependent on a single parameter $a(t)$. In the spherically symmetric case, with $\nu ,\lambda$ as two arbitrary parameters this method does \emph{not} work. Further in the first case people assumes the Deser-Boulware solution in the bulk, but in the effective equation formalism there is no need to have knowledge about the bulk metric. Hence in the situations we know the bulk metric, use of Gauss-Codazzi equation is redundant, however in our case we are \emph{not} bothered by the bulk metric and in this case there is no way to avoid Gauss-Codazzi equation.

Returning back to our discussion, it turns out that even with spherical symmetries, solution to the effective equations cannot be obtained in closed form. This has to do with the fact that the left hand side of \ref{GB_05} contains curvature tensor squared expressions. In order to obtain analytic closed form solution, on top of these symmetries we keep terms linear in the \gb coupling parameter $\alpha$. This immediately suggests that all the quadratic terms are already multiplied by $\alpha$ and thus all the quadratic terms will be replaced by their \gr\ expressions. Finally, for vacuum, spherically symmetric, static spacetime the effective field equations linear in \gb coupling constant takes the following form:
\begin{align}\label{GB_11}
G_{\mu \nu}+E_{\mu \nu}+\alpha \left\lbrace \frac{4}{3}R_{\mu \alpha \beta \gamma}R_{\nu}^{~\alpha \beta \gamma}-\frac{7}{12}h_{\mu \nu}\left(R_{\alpha \beta \gamma \delta}R^{\alpha \beta \gamma \delta}\right)-4R_{\mu \rho \nu \sigma}E^{\rho \sigma}\right\rbrace=0
\end{align}
Here $E_{\mu \nu}$ is obtained by projecting the electric part of bulk Weyl tensor on the brane, $R_{\mu \nu \alpha \beta}$ stands for four-dimensional curvature components evaluated for the background \EH action and $\alpha$ is the \gb coupling parameter. We need to solve the two unknown parameters $\lambda$ and $\nu$, for which we require two differential equations which are supplied by the temporal part and radial part of \ref{GB_11} respectively (for explicit expressions see \ref{App_E05} and \ref{GB_13} in \ref{GBApp_0102}).

The only remaining object which we need to consider corresponds to the electric part of the Weyl tensor $E_{\mu \nu} =e^{a}_{\mu}n^{b}e^{c}_{\nu}n^{d}C_{abcd}$. It is evident from the expressions that this tensor originates from the bulk Weyl tensor $C_{abcd}$. Thus it can inherit nonlocal effects from free bulk gravitational field. In this context we would like to clarify an important issue related to bulk-brane dynamics. The brane energy momentum tensor produces effects on the bulk, which in turn alters the geometry of the brane. Thus bulk and brane dynamics are entangled to each other. This effects shows up in the effective equation formalism through $E_{\mu \nu}$, as it depends on bulk Weyl tensor. In order to get the bulk Weyl tensor one needs to solve the bulk equations to relate bulk and brane dynamics, which in general is difficult to solve. However in this particular situation, when spacetime is static and spherically symmetric it is possible to encode all the bulk informations 
in $E_{\mu \nu}$ in terms of two unknown functions $U(r)$, known as ``dark radiation'' and $P(r)$, known as ``dark pressure'' (see \cite{Maartens:2001jx} for a detailed discussion). The differential equation satisfied by $U(r)$ and $P(r)$, will emerge from Bianchi identities.  

The dark radiation term $U(r)$, which is a scalar, can be obtained by projecting $E_{\mu \nu}$ along static observers world line and is given by, $U(r)=-(k_{4}/k_{5})^{4}E_{\mu \nu} u^{\mu}u^{\nu}$. Here $k_{4}^{2}=8\pi G_{N}$, with $G_{N}$ being the four-dimensional gravitational constant. Similarly, one can obtain the dark pressure as, $P_{\mu \nu}=P(r)\left(r_{\mu}r_{\nu}-\frac{1}{3}\xi _{\mu \nu} \right)$, where $r_{\mu}$ stands for the unit radial vector and $\xi _{\mu \nu}=h_{\mu \nu}+u_{\mu}u_{\nu}$ being the induced metric on $t=\textrm{constant}$ surface. The tensor $P_{\mu \nu}$ on the other hand is related to $E_{\mu \nu}$ as, $P_{\mu \nu}=-(k_{4}/k_{5})^{4}\lbrace \xi ^{\alpha}_{(\mu}\xi ^{\beta}_{\nu )}-\frac{1}{3}h_{\mu \nu}h^{\alpha \beta}\rbrace E_{\alpha \beta}$. Hence in a static spherically symmetric spacetime, $E_{\mu \nu}$ involves only two functions, the dark radiation $U(r)$ and the dark pressure $P(r)$. 

Let us now resume our main job, i.e., to find a static, spherically symmetric vacuum solution to the effective field equations presented in \ref{GB_05}. For that we need a background \gr\ vacuum solution. This was obtained in \cite{Dadhich:2000am} and the metric elements in the light of \ref{GB_10} takes the form,
\begin{align}\label{GB_15}
e^{\nu}=e^{-\lambda}=1-\frac{2GM+Q_{0}}{r}-\left(\frac{3P_{0}}{8\pi G\lambda _{T}}\right)\frac{1}{r^{2}}\equiv 1-\frac{a}{r}-\frac{b}{r^{2}}
\end{align}
where $Q_{0}$ is an integration constant and is related to the total dark radiation within a spherical volume and $\lambda_{T}$ stands for the brane tension. The two unknown functions $U(r)$ and $P(r)$ has the following functional behavior: $U(r)=-(P_{0}/2r^{4})$ and $P(r)=(P_{0}/r^{4})$, which brings $P_{0}$ in \ref{GB_15}. The two constants $a$ and $b$ are determined from the \gr\ solution and have respective values: $a=2GM+Q_{0}$ and $b=(3P_{0}/8\pi G\lambda _{T})$. We now compute the components of curvature tensor using the background \gr\ solution discussed earlier and obtain an expression for the right hand side of effective field equations as presented in \ref{App_E05} and \ref{GB_13}. This will lead to first order linear differential equations in $\nu$ and $\lambda$ correct up to linear order in \gb coupling constant. Thus performing this procedure (lengthy but straightforward; see \ref{GBAPP_01}) we arrive at the following equation for the temporal component (for a detailed derivation 
see \ref{App_E06} in \ref{GBApp_0102}),
\begin{align}\label{GB_16}
G^{t}_{t}&+3\bar{\kappa}U(r)-\alpha \left(3\frac{a^{2}}{r^{6}}+\frac{20}{3}\frac{ab}{r^{7}}+\frac{10}{3}\frac{b^{2}}{r^{8}}\right)-4\alpha \Big[-\bar{\kappa}\left(\frac{a}{r^{3}}+\frac{3b}{r^{4}}\right)\left(U+2P\right)
\nonumber
\\
&+\bar{\kappa}\frac{1}{2r}\left(U-P\right)\left(\frac{a}{r^{2}}+\frac{2b}{r^{3}}\right)+\bar{\kappa}\frac{1}{2r}\left(U-P\right)\left(\frac{a}{r^{2}}+\frac{2b}{r^{3}}\right)\Big]=0
\end{align}
where the constant $\bar{\kappa}=(1/3)(k_{5}/k_{4})^{4}=(1/4\pi G \lambda _{T})$ \cite{Chakraborty:2015wma} is related to the coefficient of electric part of Weyl tensor. In the above expression we also have the following solutions for the dark radiation $U=-P_{0}/2r^{4}$ and dark pressure $P=P_{0}/r^{4}$. These results upon substitution in \ref{GB_16} leads to
\begin{align}\label{GB_17}
G^{t}_{t}=e^{-\lambda}\left(\frac{1}{r^{2}}-\frac{\lambda '}{r}\right)-\frac{1}{r^{2}}
=\frac{1}{r^{2}}\dfrac{d}{dr}\left(re^{-\lambda}\right)-\frac{1}{r^{2}}
\end{align}
This immediately implies the following differential equation for $e^{-\lambda}$,
\begin{align}\label{GB_18}
\dfrac{d}{dr}\left(re^{-\lambda}\right)=1+\frac{3\bar{\kappa}P_{0}}{2r^{2}}-\frac{6\alpha \bar{\kappa}P_{0}}{r^{2}}\left(\frac{2a}{r^{3}}+\frac{5b}{r^{4}}\right)
+\alpha \left(3\frac{a^{2}}{r^{4}}+\frac{20}{3}\frac{ab}{r^{5}}+\frac{10}{3}\frac{b^{2}}{r^{6}}\right)
\end{align}
The left hand side is a linear differential operator acting on $e^{-\lambda}$, the $g_{rr}$ metric component and the right hand is solely a function of $r$, due to spherical symmetry. This can be readily integrated and finally we obtain:
\begin{align}\label{GB_19}
e^{-\lambda}&=1-\frac{a}{r}-\frac{b}{r^{2}}
-\alpha \left(\frac{a^{2}}{r^{4}}+\frac{5ab}{3r^{5}}+\frac{2b^{2}}{3r^{6}}-\frac{3\bar{\kappa}P_{0}a}{r^{5}}-\frac{6\bar{\kappa}P_{0}b}{r^{6}}\right)
\nonumber
\\
&=\left\lbrace 1-\frac{2GM+Q_{0}}{r}-\left(\frac{3P_{0}}{8\pi G\lambda _{T}}\right)\frac{1}{r^{2}}\right\rbrace 
\nonumber
\\
&-\alpha \left\lbrace \frac{(2GM+Q_{0})^{2}}{r^{4}}-\left(\frac{3P_{0}}{8\pi G\lambda _{T}}\right)\frac{2GM+Q_{0}}{3r^{5}}-\left(\frac{3P_{0}}{8\pi G\lambda _{T}}\right)^{2}\frac{10}{3r^{6}}\right\rbrace
\end{align}
where in obtaining the last line we have incorporated the values of $a$, $b$ and $\bar{\kappa}$ respectively, which are, $a=2GM+Q_{0}$, $b=3P_{0}/8\pi G\lambda _{T}$ and $\bar{\kappa}=(1/4\pi G \lambda _{T})$.

We note that the introduction of additional \gb correction term alters the nature of the black hole solution. The most prominent change corresponds to change in the location of the black hole horizon. In the original spacetime the horizon is located at the radius where $(1-(a/r)-(b/r^{2}))$ identically vanishes. However in the modified metric the above horizon is no longer the radius at which $f(r)$ vanishes. In the modified metric the solution of $f(r)=0$ corresponds to a sixth degree algebraic equation, whose solution can considerably differ from the original location of the black hole horizon. The difference should depend linearly on $\alpha$ since our metric elements are valid only within linear orders in the \gb coupling parameter. Hence if the original horizon was located at $r=r_{0}$, the modified horizon should be expressible as, $r_{h}=r_{0}+\alpha r_{1}+\mathcal{O}(r^{2})$. Here the parameter $r_{1}$ depends on the mass, $Q_{0}$ and $P_{0}$, inherited from the bulk. Moreover for consistency of the 
perturbative method this term should be smaller compared to the background value. 

Along similar lines we can use the background \gr\ metric elements to compute the right hand side of \ref{GB_13}. Then it will again provide a linear differential equation in $e^{\nu}$. Since due to spherical symmetry the right hand side of \ref{GB_13} depends only on the radial coordinate $r$, that can be readily integrated leading to solution for $e^{\nu}$. The differential equation for $e^{\nu}$ originating from $G^{r}_{r}$ turns out to be (for a detailed derivation see \ref{App_E07} of \ref{GBApp_0102}),
\begin{align}\label{GB_20}
e^{-\lambda}\Bigg(\frac{\nu '}{r}&+\frac{1}{r^{2}}\Bigg)-\frac{1}{r^{2}}=\left(\frac{3P_{0}}{8\pi G \lambda _{T}}\right)\frac{1}{r^{4}}+\alpha \Bigg[ \Bigg\lbrace 3\frac{(2GM+Q_{0})^{2}}{r^{6}}+\frac{5(2GM+Q_{0})P_{0}}{2\pi G\lambda _{T}}\frac{1}{r^{7}}
\nonumber
\\
&+\frac{10}{3}\left(\frac{3P_{0}}{8\pi G\lambda _{T}}\right)^{2}\frac{1}{r^{8}}\Bigg\rbrace -\frac{3P_{0}}{2\pi G\lambda _{T}}\frac{1}{r^{4}}\Bigg\lbrace 2\frac{2GM+Q_{0}}{r^{3}}+5\left(\frac{3P_{0}}{8\pi G\lambda _{T}} \right)\frac{1}{r^{4}}\Bigg\rbrace \Bigg]
\end{align}
Then finally integrating this expression we readily obtain the solution for $g_{tt}$ component as $e^{\nu}=e^{-\lambda}$, where $e^{-\lambda}$ is given by \ref{GB_19}. In the background spacetime described by \gr\ we had $g_{tt}=g_{rr}^{-1}$, then it turns out that even when \gb correction term is added to the \EH Lagrangian, to first order in \gb coupling parameter, the vacuum solution still satisfies the condition $g_{tt}=g_{rr}^{-1}$.

\begin{figure*}
\begin{center}
\includegraphics[scale=1]{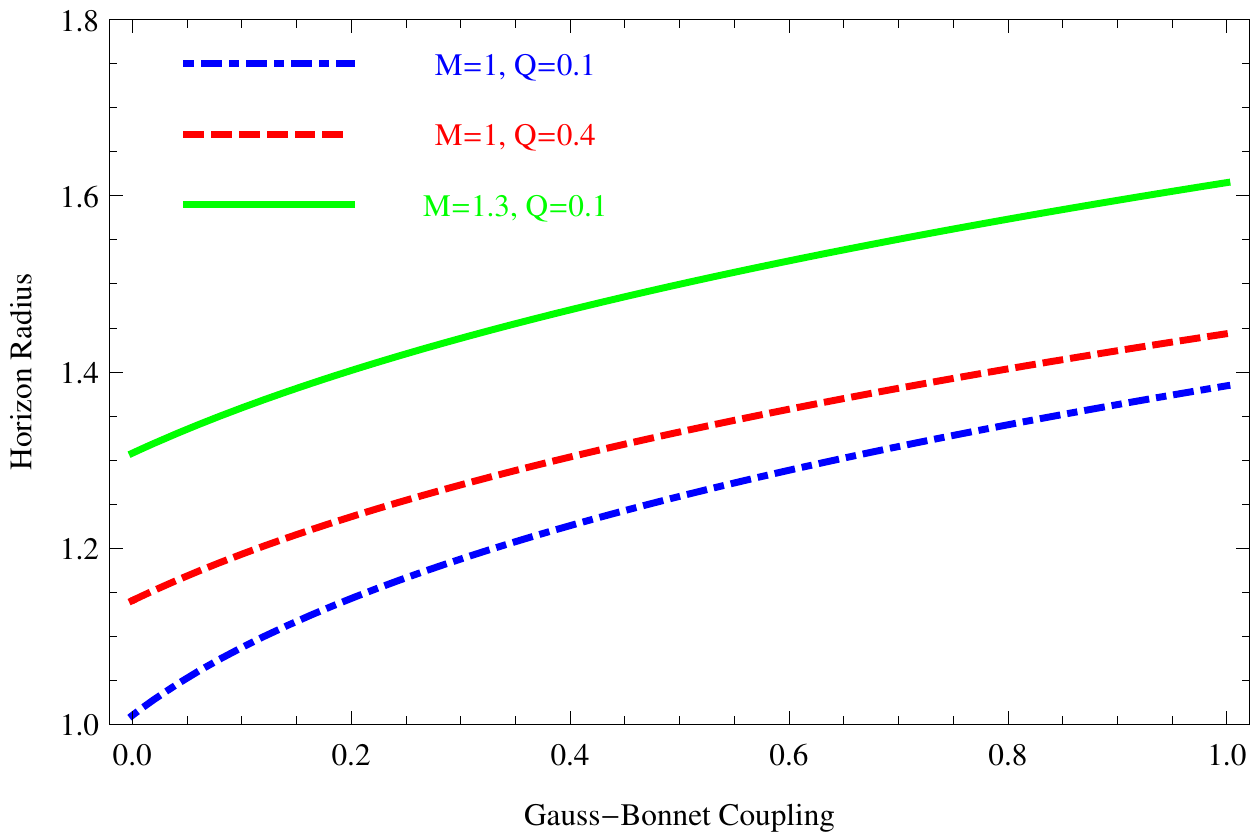}

\caption{The figure illustrates variation of the horizon radius with the \gb coupling constant $\alpha$. The coefficient of $r^{-1}$ is denoted as $M$, while that for $r^{-2}$ is $Q^{2}$. Hence $M$ is the standard gravitational mass, while $Q$ is the gravitational charge inherited from the bulk. As expected, with increase of $M$ and $Q$ the horizon radius increases, as well as with increase of Gauss-Bonnet coupling constant. The unit has been chosen such that $G=1=c$, for convenience.}
\label{fig_01}
\end{center}
\end{figure*}

Note that the above analysis has been carried away to the first order in the \gb coupling parameter. At this stage it might seem a subjective choice but as we will numerically explain a few paragraphs down the line this is a justified procedure. Higher order contributions are much smaller. However given the solution at linear order one might draw some tentative for of the metric function at second order. For example, it seems natural that the second order contributions should contain, $\alpha ^{2}(2GM+Q_{0})^{4})r^{-8}$, $\alpha ^{2}(3P_{0}/8\pi G\lambda _{T})^{2}(2GM+Q_{0})^{2})r^{-10}$, $\alpha ^{2}(3P_{0}/8\pi G\lambda _{T})^{4}r^{-12}$ and so on. As emphasized earlier these terms would be very small in comparison with the first order terms, which we will discuss later.
\paragraph*{Physical Relevance of the Solution} The solution obtained above inherits the effect from the bulk spacetime as well as from the \gb term, we will discuss both the terms separately. The bulk, affects the brane through the electric part of the Weyl tensor which manifests itself through the dark pressure term $P_{0}$ and the brane tension $\lambda _{T}$. Note that the corresponding term in the metric, $(3P_{0}/8\pi G\lambda _{T})r^{-2}$ has appearance as the electric charge but with a negative sign. This originates due the fact that the ``charge'' is gravitational in nature, generated from the bulk dynamics. The horizon location corresponds to $e^{-\lambda}=0$, which without the \gb coupling leads to,
\begin{equation}
r_{\rm h,0}=\frac{1}{2}\left[\left(2GM+Q_{0}\right)\pm \sqrt{\left(2GM+Q_{0}\right)^{2}+\frac{3P_{0}}{2\pi G\lambda _{T}}}\right]
\end{equation}
As the \gb coupling is included, the horizon location gets modified to linear order in $\alpha$ as,
\begin{equation}
r_{\rm h}=r_{\rm h,0}+\alpha \left(\frac{\left(2GM+Q_{0}\right)^{2}}{4r_{\rm h,0}^{3}-3\left(2GM+Q_{0}\right)r_{\rm h,0}^{2}
-\left(\frac{3P_{0}}{4\pi G\lambda _{T}}\right)r_{\rm h,0}} \right)
\end{equation}
In the general case the closed form expression is quiet involved, so we have presented the variation of the horizon radius with the \gb coupling in \ref{fig_01}. It turns out that the horizon radius increases with the increase of the \gb coupling parameter.

Two more aspects can be briefly mentioned in this connection. The first one corresponds to the horizon thermodynamics. One can invert the above relation to obtain gravitational mass in terms of the horizon radius and can further be expressed in terms of the Wald entropy suited for Einstein-Gauss-Bonnet gravity, which leads to the black hole temperature. One can calculate the corresponding specific heat straightforwardly, leading to standard and finite behavior. As expected, the above solution does not exhibit any phase transition. The other aspect, corresponds to stability of the solution. This can be performed by considering the linear perturbations over and above the metric functions. The perturbations can be divided into three classes --- scalar, vector and tensor. All these modes satisfy decoupled wave equations with a potential depending on the metric functions. For stability the potentials should be positive \cite{Kodama:2003kk,Ishibashi:2003ap,Kodama:2003jz}. Since all the dominating terms, the bulk 
gravitational charge and 
the \gb correction term appears with a 
negative sign, their contribution to the potentials for scalar and tensor modes would be positive, while that for the vector modes would be negative as one can readily verify. Hence under linear tensor and scalar perturbations, the above solutions are stable.
\paragraph*{Numerical Estimations} To understand the approximations involved, it is instructive to provide a numerical estimate of curvature scales associated with the problem. The potential for \gr\ scales as $R_{S}/r$, where $R_{S}$ is the Schwarzschild radius. For sun, one has $R_{S}=2.95~ \textrm{km}$. On the other hand from solar system physics, the \gb parameter is constrained to $\alpha \lesssim 1.27\times 10^{-7}$ \cite{Chakraborty:2012sd}. Hence the ratio of the curvature due to first order correction compared to the \gr\ solution corresponds to: $(\alpha R_{S})/r^{3}$, where $r$ is the distance from the singularity. For $r=\eta R_{S}$, where $\eta$ is a numerical factor one obtains the ratio to be $0.15~\eta ^{-3}\times 10^{-13}$, for a solar mass black hole. Hence it suffices to keep terms linear in $\alpha$, as long as one is interested in energy scales much below Planck scale (or, away from the singularity). This justifies our expansion of the field equations on brane to linear 
order in the \gb coupling parameter.

Having obtained the vacuum spherically symmetric static solution for \EH action with \gb correction term to linear order in the \gb coupling parameter, we will now take up the case of $f(\mathcal{R})$ gravity added to \gb correction term and shall obtain the vacuum solution thereof.
\section{Vacuum Solution in the brane with bulk f(R) and Gauss-Bonnet Gravity}\label{frgb}

We will now concentrate on the situation when both $f(\mathcal{R})$ and \gb gravity are present in the bulk, such that the bulk gravitational action takes the form presented in \ref{GB_06}. With vacuum, static, spherically symmetric spacetime the line element can be written as \ref{GB_10}, with $g_{tt}=e^{\nu(r)}$, $g_{rr}=e^{\lambda(r)}$.

In this case also, these symmetries cannot help to analytically solve the problem and we once again work upto linear order in $\alpha$, the \gb coupling constant. Using the symmetries and the above condition we arrive at the following form for the effective field equations,
\begin{align}\label{GB_FR01}
G_{\mu \nu}+E_{\mu \nu}-F(\mathcal{R})h_{\mu \nu}=-\alpha \left\lbrace \frac{4}{3}R_{\mu \alpha \beta \gamma}R_{\nu}^{~\alpha \beta \gamma}-\frac{7}{12}h_{\mu \nu}\left(R_{\alpha \beta \gamma \delta}R^{\alpha \beta \gamma \delta}\right)-4R_{\mu \rho \nu \sigma}E^{\rho \sigma}\right\rbrace
\end{align}
where again $E_{\mu \nu}$ stands for the electric part of the bulk Weyl tensor. This is the part which originates from bulk and in general we need to solve bulk equations in order to get $E_{\mu \nu}$, which is a formidable task. As emphasized earlier in presence of staticity and spherical symmetry $E_{\mu \nu}$ can be decomposed into a dark radiation part and a dark pressure part as in \ref{GB_SphBrane}. Given this setup we shall now be able to determine the solution for the two unknown parameters, namely $\lambda$ and $\nu$. For the background metric we need solution to the effective field equations in $f(\mathcal{R})$ gravity model. Such a solution was obtained in \cite{Chakraborty:2014xla} and corresponds to the following line element,
\begin{align}\label{GB_FR02}
ds^{2}=1-\frac{2GM+Q_{0}}{r}-\left(\frac{3P_{0}}{8\pi G\lambda _{T}}\right)\frac{1}{r^{2}}+\frac{F(\mathcal{R})}{3}r^{2}\equiv 1-\frac{a}{r}-\frac{b}{r^{2}}+cr^{2}
\end{align}
Here $Q_{0}$ stands for an integration constant and is related to the total dark radiation within a spherical volume. $\lambda_{T}$ denotes the brane tension. The origin of $P_{0}$ in \ref{GB_FR02} is from the dark radiation term i.e., from the bulk Weyl tensor. The three constants $a$, $b$ and $c$ are determined from the $f(\mathcal{R})$ solution to have the respective values: $a=2GM+Q_{0}$, $b=(3P_{0}/8\pi G\lambda _{T})$ and $c=F(\mathcal{R})/3$. Having obtained the metric it is straightforward to compute the components of curvature tensor and obtain an expression for the right hand side of the effective field equations as presented in \ref{GB_FR01}. This will eventually lead to first order linear differential equations in $\nu$ and $\lambda$ correct up to linear order in \gb coupling constant. Performing this procedure (see \ref{GBAPP_01}) we arrive at the following differential equation for $e^{-\lambda}$ (see \ref{App_E08} in \ref{GBApp_0102} for a detailed derivation),
\begin{align}\label{GB_FR03}
\dfrac{d}{dr}(re^{-\lambda})&=1+\frac{3\bar{\kappa}P_{0}}{2r^{2}}+\alpha \left(3\frac{a^{2}}{r^{4}}+\frac{20}{3}\frac{ab}{r^{5}}+\frac{10}{3}\frac{b^{2}}{r^{6}}+\frac{20}{3}c^{2}r^{2}+\frac{16}{3}\frac{bc}{r^{2}}\right)
\nonumber
\\
&+F(\mathcal{R})r^{2}-\frac{6\alpha \bar{\kappa}P_{0}}{r^{2}}\left(\frac{2a}{r^{3}}+\frac{5b}{r^{4}}+c\right)
\end{align}
where we have introduced this quantity $\bar{\kappa}=(1/4\pi G\lambda _{T})$. The above equation of $e^{-\lambda}$, being an ordinary, first order differential equation can be readily integrated and thus a solution for $e^{-\lambda}$ can be obtained. This solution turns out to be,
\begin{align}\label{GB_FR04}
e^{-\lambda}&=1-\frac{2GM+Q_{0}}{r}-\frac{3\bar{\kappa}P_{0}}{2r^{2}}+F(\mathcal{R})\frac{r^{2}}{3}
-\alpha \left(\frac{a^{2}}{r^{4}}-\frac{ab}{3r^{5}}-\frac{10b^{2}}{3r^{6}}-\frac{20}{9}c^{2}r^{2}+\frac{4}{3}\frac{bc}{r^{2}}\right)
\nonumber
\\
&=\left(1-\frac{2GM+Q_{0}}{r}-\frac{3\bar{\kappa}P_{0}}{2r^{2}}+F(\mathcal{R})\frac{r^{2}}{3}\right)
\nonumber
\\
&-\alpha \Bigg[\frac{(2GM+Q_{0})^{2}}{r^{4}}-\left(\frac{3P_{0}}{8\pi G\lambda _{T}}\right)\frac{2GM+Q_{0}}{3r^{5}}-\left(\frac{3P_{0}}{8\pi G\lambda _{T}}\right)^{2}\frac{10}{3r^{6}}
-\frac{20}{81}F(\mathcal{R})^{2}r^{2}+\frac{4}{3}\frac{P_{0}F(\mathcal{R})}{8\pi G\lambda _{T}r^{2}}\Bigg]
\end{align}
where in order to obtain the last line we have substituted for $a=2GM+Q_{0}$, $b=(3P_{0}/8\pi G\lambda _{T})$ and $c=F(\mathcal{R})/3$. Also note that for $e^{-\lambda}$ the first four terms constitute the standard $f(\mathcal{R})$ gravity solution \cite{Chakraborty:2014xla}, plus additional correction terms of order $\alpha$, the \gb coupling parameter, as the action is modified by the addition of a \gb term. 

In obtaining this solution we have used the fact that $F(\mathcal{R})$ on the brane is merely a constant, which in turn implies that in the scale-tensor representation the scalar field is a constant on the brane. Since the scalar field does not vanish at infinity, the no-hair theorems are not applicable and hence the spherically symmetric solution can differ from its general relativistic counterpart. This difference is manifested in the term $F(\mathcal{R})(r^{2}/3)$. Similar, non-trivial solutions were obtained earlier in \cite{Capozziello:2007wc,Multamaki:2006zb,Multamaki:2006ym,Borzou:2009gn,Haghani:2013oma,Carames:2012gr,
Chakraborty:2014xla,Chakraborty:2015bja}. Further for $F(\mathcal{R})<0$, one obtains Schwarzschild de Sitter solution on the brane when \gb parameter vanishes. As the \gb parameter appears the structure differs from that of Schwarzschild de Sitter, which was also noted earlier in \cite{delaCruzDombriz:2009et}.

We note that due to this addition of \gb correction term to the original $f(\mathcal{R})$ gravity action the location of the black hole horizon changes. A black hole solution in the original spacetime can be altered to a different location. This is evident, since the original fourth order algebraic equation gets transformed to a sixth order one. Depending on the parameter space of the parameters, consistent with the expansion to linear order in \gb parameter, the black hole horizon will be modified by a term linear in $\alpha$.

\begin{figure*}
\begin{center} 
\includegraphics[scale=0.6]{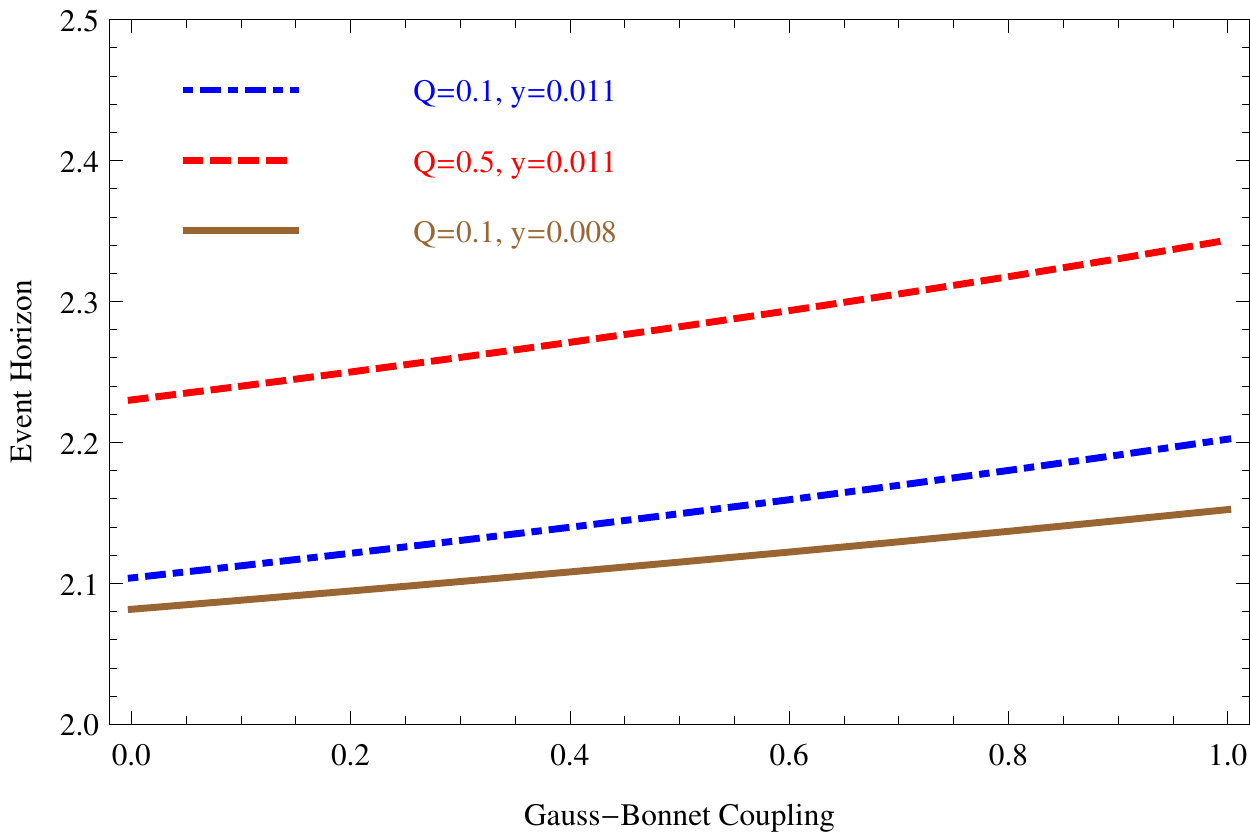}~~
\includegraphics[scale=0.6]{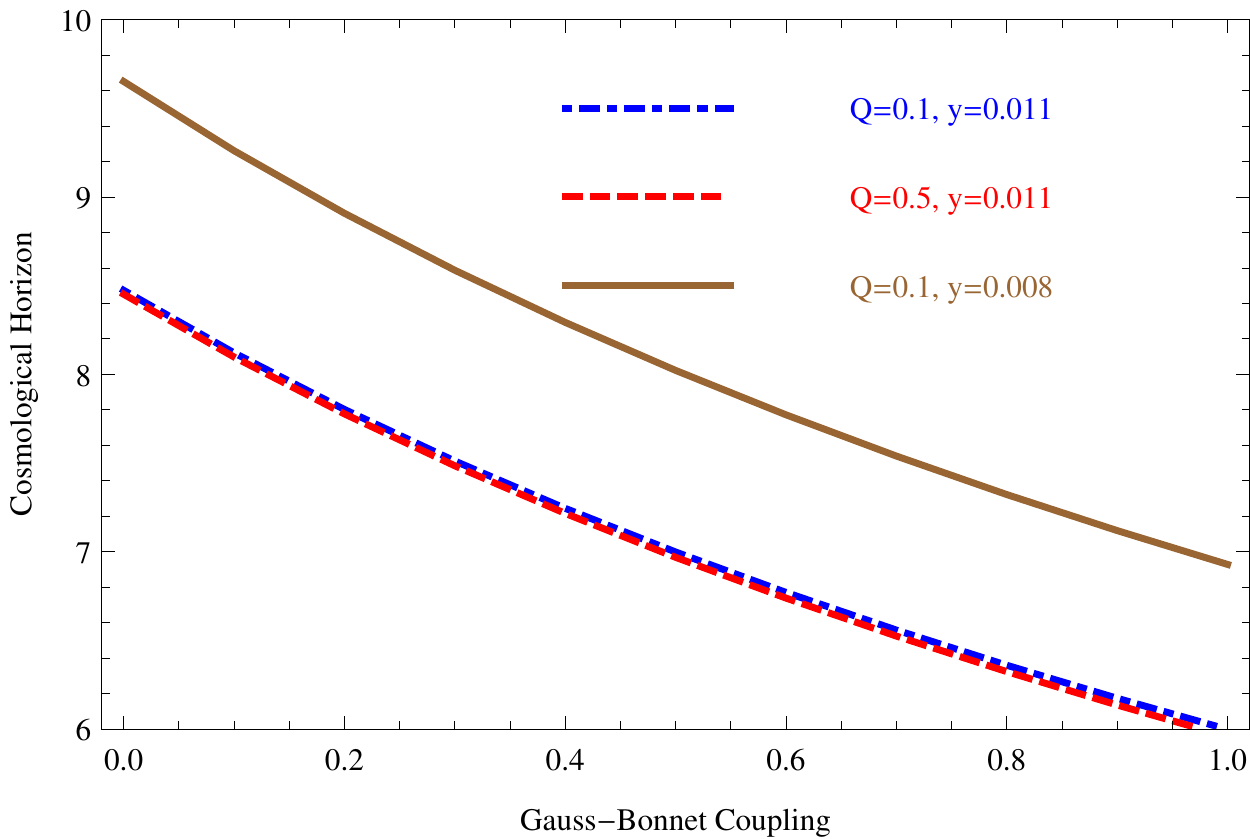}

\caption{The figure illustrates variation of both the event and cosmological horizon with the \gb coupling constant $\alpha$. The variation with the gravitational charge inherited from the bulk and the contribution from $f(\mathcal{R})$ gravity is also presented. The event horizon radius increases with increase of gravitational charge $Q$, while it decreases with decrease of $F(\mathcal{R})$. While for the cosmological horizon, the effect of increasing (or decreasing) the gravitational charge $Q$ is negligible, but as the term $F(\mathcal{R})$ is decreased, the cosmological horizon radius increases. The unit has been chosen such that $G=1=c$ and $M=1$, for convenience.}
\label{fig_02}
\end{center}
\end{figure*}

Now we have to solve for $e^{\nu}$ in order to obtain the full spherically symmetric solution. For that we require the radial part of the effective equations. Like $e^{\lambda}$, for $e^{\nu}$ as well we have first order and ordinary differential equation. This differential equations turns out to have the following form (see \ref{App_E10} in \ref{GBApp_0103} for detailed discussion), 
\begin{align}\label{GB_FR05}
e^{-\lambda}\Bigg(\frac{\nu '}{r}&+\frac{1}{r^{2}}\Bigg)-\frac{1}{r^{2}}=F(\mathcal{R})+\frac{3P_{0}}{8\pi G\lambda _{T}}\frac{1}{r^{4}}+\alpha \Bigg[\Bigg\lbrace \frac{3\left(2GM+Q_{0}\right)^{2}}{r^{6}}+\frac{10}{3}\left(\frac{3P_{0}}{8\pi G\lambda _{T}}\right)^{2}\frac{1}{r^{8}}
\nonumber
\\
&+\frac{20}{27}F(\mathcal{R})^{2}+\frac{20}{3}\left(\frac{3P_{0}}{8\pi G\lambda _{T}}\right)\frac{2GM+Q_{0}}{r^{7}}+\frac{10}{9}\left(\frac{3P_{0}}{8\pi G\lambda _{T}}\right)\frac{F(\mathcal{R})}{r^{4}}\Bigg\rbrace 
\nonumber
\\
&-\frac{3P_{0}}{2\pi G\lambda _{T}}\frac{1}{r^{4}}\Bigg\lbrace \frac{2\left(2GM+Q_{0}\right)}{r^{3}}+\left(\frac{3P_{0}}{8\pi G\lambda _{T}}\right)\frac{5}{r^{4}}+\frac{F(\mathcal{R}}{3})\Bigg\rbrace \Bigg]
\end{align}
This equation can be integrated and we readily obtain the solution for $g_{tt}$ component as $e^{\nu}=e^{-\lambda}$, where $e^{-\lambda}$ is given by \ref{GB_FR04}. In the background spacetime described by $f(\mathcal{R})$ gravity we had $g_{tt}=g_{rr}^{-1}$. To our surprise, it turns out that even when \gb correction term is added to the $f(\mathcal{R})$ Lagrangian, to first order in \gb coupling parameter the vacuum solution still satisfies the condition $g_{tt}=g_{rr}^{-1}$.
\paragraph*{Physical Relevance of the Solution} To complete the discussion, let us consider the physical relevance of the solution derived above. As in the previous situation, here also the horizon location will change. However for $F(\mathcal{R})<0$, the solution being of the form Schwarzschild de Sitter, one expects two horizons --- the black hole event horizon and the cosmological event horizon. The discussion is simpler in units where $M=1$. Thus for $\alpha =0$, one defines, dimensionless parameters, $Q^{2}=(3\bar{\kappa}P_{0}/2M^{2})$ and $y=-[F(\mathcal{R})/3]M^{2}$, such that the location of the horizons, $r_{\rm h,0}$ can be obtained by solving the following algebraic relation,
\begin{equation}
1-\frac{2}{r_{\rm h,0}}-\frac{Q^{2}}{r_{\rm h,0}^{2}}-yr_{\rm h,0}^{2}=0
\end{equation}
Due to introduction of the \gb coupling, the location of both cosmological and black hole event horizon with change. The modification to linear order in \gb coupling constant yield,
\begin{equation}
r_{\rm h}=r_{\rm h,0}+\bar{\alpha}\left(\frac{\frac{20}{9}y^{2}r_{\rm h,0}^{4}+\frac{4}{3}Q^{2}y}{4yr_{\rm h,0}^{3}+6r_{\rm h,0}^{2}-2r_{\rm h,0}}\right) 
\end{equation}
where, $\bar{\alpha}$ is another dimensionless parameter, $(\alpha/M^{2})$. The general expression for modified horizon locations is quiet involved, so we have computed it numerically and presented the variation of the horizons with \gb coupling parameter in \ref{fig_02}. Note this this shift in the horizon location is quiet crucial from experimental point of view. Regarding the displacement of event horizon, have implications on the accretion disk structure, gravity wave emission, which might be detectable in near future with more and more accurate determination of the event horizon. On the other hand, modification in cosmological horizon can also act as a test bed for \gb theories along with $f(\mathcal{R})$ gravity. Since typically the scale below which galaxy clusters can exist corresponds to the cosmological horizon, hence any shift in the cosmological horizon might have observable effect on the structure formation in the universe.

Let us now briefly comment on the thermodynamic behavior and stability of the solution. Using the equation $e^{-\lambda}=0$, we can obtain the gravitational mass as a function of the horizon radius and subsequently use of entropy corresponding to $f(\mathcal{R})$ with \gb gravity leads to black hole temperature in terms of the entropy. Taking derivative of the temperature one immediately obtains the specific heat. For $F(\mathcal{R})<0$, the specific heat turns out to be finite, while for $F(\mathcal{R})>0$, the specific heat diverges and changes sign, signaling appearance of a second order phase transition (for details see \cite{Chakraborty:2014xla}). Moreover both the scalar and tensor modes are found to be stable for any possible values of the parameters, while for $F(\mathcal{R})>0$, the vector mode would be stable, otherwise it leads to instability \cite{Chakraborty:2014xla}. Similar situation is found for the Einstein-Gauss-Bonnet black hole as well.

Further, the numerical estimation of various parameters scale exactly like the previous scenario, showing the validity of expanding the field equations to linear order in \gb coupling parameter. So far we have obtained the vacuum spherically symmetric static solution for both the \EH action and $f(\mathcal{R})$ action with \gb correction term. We will now take up the situation where field equations on the brane are obtained by perturbative method and shall comment on the spherically symmetric solutions obtained from them. 
\section{Spherically symmetric solution on the brane using perturbative method}\label{pergb}

In the previous sections we have derived the spherically symmetric solutions on the brane starting from the field equations obtained by projecting the bulk equations using Gauss-Codazzi relations. As emphasized earlier, the other perturbative method corresponding to expansion of bulk equations in terms of bulk to brane curvature ratio. Due to fundamental difference between these two approaches (one is exact, while the other one is perturbative) in this section we will briefly describe how to obtain static spherically symmetric solution to these field equations derived assuming $\textrm{bulk~curvature~scale}\ll\textrm{brane~curvature~scale}$. 
\paragraph*{Brane solution corresponding to Einstein-Gauss-Bonnet bulk}

For \EH action with \gb correction term, the perturbative field equations have been presented in \ref{GB_09}. The right hand side of \ref{GB_09} in absence of external matter fields and for spherically symmetric spacetime becomes linear in the \gb coupling parameter, simplifying our analysis. Further expanding the equations to linear order in \gb coupling parameter, using \ref{GB_15} as the background spacetime we finally obtain the differential equation for $e^{-\lambda}$ to be (see \ref{App_E11} in \ref{GBApp_0104} for derivation),
\begin{align}
\dfrac{d}{dr}\left(re^{-\lambda}\right)&=1+\frac{3\bar{\kappa}P_{0}}{2r^{2}}+\frac{4\beta}{\ell (1+\beta)}~^{(2)}\chi ^{t}_{t}(r)~r^{2}
\nonumber
\\
&-\frac{\beta \ell ^{2}}{12}\left(\frac{3}{2}\frac{a^{2}}{r^{4}}+\frac{8ab}{r^{5}}+\frac{11b^{2}}{r^{6}}\right)
+\frac{7\beta \ell ^{2}}{12}\left(\frac{3a^{2}}{2r^{4}}+\frac{4ab}{r^{5}}+\frac{3b^{2}}{r^{6}}\right)
\end{align}
where $a=2GM+Q_{0}$, $b=(3P_{0}/8\pi G\lambda _{T})$. The most important object appearing in the above equation is $^{(2)}\chi ^{t}_{t}$, which originates as we integrate the bulk equations over extra coordinate and thus in spherically symmetric case it is dependent on the radial coordinate only. This equation can be readily integrated leading to the following expression for $e^{-\lambda}$ (see \ref{App_E12} in \ref{GBApp_0104} for derivation),
\begin{align}
e^{-\lambda}=1-\frac{2GM+Q_{0}}{r}-\frac{3\bar{\kappa}P_{0}}{2r^{2}}+\frac{16 \alpha}{\ell ^{3}}\frac{1}{r}\int ~^{(2)}\chi ^{t}_{t}(r)~r^{2}dr-\frac{\alpha}{3}\left(3\frac{a^{2}}{r^{4}}+\frac{5ab}{r^{5}}+\frac{2b^{2}}{r^{6}}\right)
\end{align}
where we have written $\beta$ in terms of the \gb coupling parameter $\alpha$ and have defined $\bar{\kappa}=(1/4\pi G\lambda _{T})$. It is clear that for $\alpha =0$, we retrieve the standard solution of \gr. All the correction terms are linear in $\alpha$ and originates from the addition of \gb term to the original \EH action. However still this solution has some arbitrariness, in the sense that the object $^{(2)}\chi ^{t}_{t}$ has not been determined. The most remarkable feature of this solution comes into light if we choose the bulk integration constant $^{(2)}\chi ^{t}_{t}$ to be,
\begin{align}\label{chichoice}
^{(2)}\chi ^{t}_{t}=\ell ^{3}\left(-\frac{ab}{2r^{7}}-\frac{5b^{2}}{4r^{8}}\right)
\end{align}
In which case the solution for $e^{-\lambda}$ can be obtained exactly by substituting for $a=2GM+Q_{0}$, $b=(3P_{0}/8\pi G\lambda _{T})$ and $\bar{\kappa}=(1/4\pi G\lambda _{T})$, which matches \emph{exactly} with the solution obtained from the projected field equations on the brane. One further obtains $e^{\nu}=e^{-\lambda}$. Note that, in the first approach we have expanded the brane field equations to linear order in \gb parameter, while in the second approach, we have also assumed $\textrm{bulk~curvature~scale}\ll\textrm{brane~curvature~scale}$.
\paragraph*{Brane solution corresponding to $\boldsymbol{f(\mathcal{R})}$-Gauss-Bonnet bulk}

To complete the discussion, we will try to see whether the same result holds for $f(\mathcal{R})$ gravity with \gb correction term as well. The perturbative field equations for $f(\mathcal{R})$ gravity with \gb correction term is presented in \ref{GB_New05}. Keeping terms linear in the \gb coupling strength and using the background spacetime given in \ref{GB_FR02} the metric element $e^{-\lambda}$ becomes
\begin{align}
e^{-\lambda}&=1-\frac{2GM+Q_{0}}{r}-\frac{3P_{0}}{8\pi G\lambda _{T}}\frac{1}{r^{2}}+\frac{F(\mathcal{R})-\Lambda _{4}}{3}r^{2}+\frac{16 \alpha}{\ell ^{3}}\frac{1}{r}\int ~^{(2)}\chi ^{t}_{t}(r)~r^{2}dr
\nonumber
\\
&-\frac{\alpha}{3}\Big[\frac{3a^{2}}{r^{4}}+\frac{2b^{2}}{r^{6}}+\frac{5ab}{r^{5}}-6c^{2}r^{2}-\frac{12bc}{r^{2}}\Big]
\end{align}
Here $a=2GM+Q_{0}$, $b=(3P_{0}/8\pi G\lambda _{T})$ and $c=F(\mathcal{R})/3$ are the parameters from the background spacetime. Finally setting $^{(2)}\chi ^{t}_{t}(r)$ to the following value,
\begin{align}
^{(2)}\chi ^{t}_{t}=\ell ^{3}\left(-\frac{ab}{r^{7}}-\frac{5b^{2}}{4r^{8}}+\frac{bc}{3r^{4}}\right)
\end{align}
one one arrives at \ref{GB_FR04}. Thus we have derived the spherically symmetric solutions for the perturbative field equations in both the gravity theories as well (see \ref{GBApp_0105} for a detailed derivation). 
\paragraph*{Numerical Estimation} As the solutions derived in perturbative method matches with the solutions derived in earlier sections the numerical comparison is more or less similar. In the perturbative calculation, however, there are two scales involved. One corresponds to the curvature scale due to \gb term to the curvature scale of the background spacetime. As we have illustrated earlier, this corresponds to, $(\alpha R_{S})/r^{3}$, where $r$ is the distance from the singularity and $R_{S}$ is the Schwarzschild radius. For $r=\eta R_{S}$, $\eta$ being a numerical factor, one obtains, the ratio to yield, $0.15~\eta ^{-3}\times 10^{-13}$, for a solar mass black hole. Hence the approximation of keeping terms linear in the \gb coupling parameter is quiet justified at the curvature scales under consideration. On the other hand the other approximation,  $\textrm{bulk~curvature~scale}\ll\textrm{brane~curvature~scale}$ can also be numerically tested. The bulk curvature scale corresponds to 
Planck Length $\sim 10^{-35}~\textrm{m}$, while the brane curvature scale for a solar mass black hole near the event horizon corresponds to,  $\sim 10^{3}~\textrm{m}$. Hence the bulk curvature scale is way smaller compared to that on the brane. Thus both the approximations used to derive the solution are well satisfied.
\section{Discussion}

In the literature, there exist two possible ways to derive gravitational field equations on a brane starting from the higher dimensional bulk equations. One of them is arrived at by projecting the bulk geometrical quantities on the brane, which is accomplished through use of Gauss-Codazzi equations and their various contractions. This process inherits the bulk effect through the electric part of the Weyl curvature tensor. In static spherically symmetric spacetime this term can be expressed through two unknown scalar functions, known as dark radiation and dark pressure respectively, with an equation of state relating them.

The second one is obtained through a perturbative expansion of the bulk field equations with bulk to brane curvature scale ratio as the perturbation parameter. In this approach the bulk equations are solved iteratively, i.e., the solution for the extra dimensional part (keeping the brane metric arbitrary) obtained at the second order is substituted in the third order equation in order to get the field equations on the brane at third order. In this manner the brane gravitational field equations are obtained. At face value it seems that these two approaches are based on completely different setup and hence would lead to completely different solutions. 

In this work we embark to show that this two approaches can lead to identical results in some simplified scenarios. Since addition of a higher curvature term to the \EH action can lead to non-trivial results, we start with \EH action with a \gb correction term. From the projected effective equations we calculate corrections to the background static spherically symmetric metric to linear order in the \gb coupling parameter. These additional terms can change the original black hole solution to a new black hole solution with an altered location of the event horizon. The solution so obtained is stable under tensor and scalar perturbations, with finite specific heat. While, for the case of $f(\mathcal{R})$ and \gb gravity present in the bulk action, the presence of \gb term modifies both the black hole and the cosmological event horizon. This might have potential signatures in the large scale structure formation. In this case as well the solution is stable under tensor and scalar perturbations. However for some 
specific choices for the derived quantity $F(\mathcal{R})$, the black hole can exhibit second order phase transition. We have also provided numerical estimate for various terms appearing in the metric and justification of the approximations made. Hence we conclude that the background geometry can be significantly distorted by addition of a higher curvature term to the \EH action. More importantly, when similar exercise is performed using the perturbative approach, one ends us getting \emph{identical} solutions. To our surprise this feature continues to hold even for bulk $f(\mathcal{R})$ gravity with a \gb correction term. This shows that these two apparently independent methods of getting the field equations on the brane yield identical static, spherically symmetric solutions in the context of both \EH and $f(\mathcal{R})$ action with a \gb correction term, keeping linear order terms in the \gb coupling parameter. 

As a possible future application one might try to explicitly demonstrate the equivalence of these two approaches by solving the full equations. In other words, one can solve the full brane field equations and compare with both the --- i) projective approach using Gauss-Codazzi relations on the brane and ii) the perturbative result, respectively.
\section*{Acknowledgements}

S.C. thanks IACS, India for warm hospitality; a part of this work was completed there during a visit. He also thanks CSIR, Government of India, for providing a SPM fellowship. 
\appendix
\labelformat{section}{Appendix #1} 
\labelformat{subsection}{Appendix #1}
\section{Appendix: Detailed Derivations}\label{GBAPP_01}

In this appendix, we summarize all the detailed derivations of respective expressions presented in the main text. We hope this will be helpful to clarify all the algebraic steps necessary to arrive at definite results in the main body of this paper.
\subsection{Basic Ingredients}\label{GBApp_0101}

The effective field equations obtained through projection of bulk gravitational field equations and Gauss-Codazzi equation takes the following form:
\begin{align}\label{App_E01}
G_{\mu \nu}&+E_{\mu \nu}-KK_{\mu \nu}+K_{\mu \rho}K^{\rho}_{\nu}+\frac{1}{2}\left(K^{2}-K_{\alpha \beta}K^{\alpha \beta}\right)h_{\mu \nu}+\alpha \left(H^{(1)}_{\mu\nu}+H^{(2)}_{\mu\nu}
+H^{(3)}_{\mu\nu}\right)
\nonumber
\\
&=\frac{2\kappa _{5}^{2}}{3}\left[\left\lbrace \mathcal{T}_{ab}e_{\mu}^{a} e_{\nu}^{b}+\left(\mathcal{T}_{ab}n^{a}n^{b}-\frac{1}{4}\mathcal{T}\right)h_{\mu\nu}\right\rbrace +\frac{\alpha}{3+\alpha M}\left(M_{\mu\nu}-\frac{1}{4}Mh_{\mu\nu}\right)\mathcal{T}_{ab}h^{ab}\right]
\end{align}
where most of the tensors have been discussed in detail in the main text with explicit expressions given. However the three tensors $H^{(1)}_{\mu\nu}$, $H^{(2)}_{\mu\nu}$ and $H^{(3)}_{\mu\nu}$ have been discussed but detailed expressions were not provided in the main text, the same appears below,  
\begin{align}
H^{(1)}_{\mu\nu}&=\frac{4}{3}\left(M_{\mu\alpha\beta\gamma}M_{\nu}^{~\alpha\beta\gamma}-3M^{\rho\sigma}
M_{\mu\rho\nu\sigma}+2MM_{\mu\nu}-4M_{\mu\rho}M_{\nu}^{~\rho}\right)
\nonumber
\\
&-\frac{1}{12}h_{\mu\nu}\left(11M^2-40M_{\alpha\beta}M^{\alpha\beta}
+7M_{\alpha\beta\gamma\delta}M^{\alpha\beta\gamma\delta}\right)
\nonumber 
\\
&-\frac{\alpha}{3(3+\alpha M)}\left(M_{\mu\nu}-\frac{1}{4}Mh_{\mu\nu}\right)
\left(M^2-8M_{\alpha\beta}M^{\alpha\beta}
+M_{\alpha\beta\gamma\delta}M^{\alpha\beta\gamma\delta}\right),
\label{App_E02}
\\
H^{(2)}_{\mu\nu}&=-4\left(M_{\mu\rho}E^\rho_{~\nu}
+M_{\nu\rho}E^\rho_{~\mu}+
M_{\mu\rho\nu\sigma}E^{\rho\sigma}
\right)+3h_{\mu\nu}M_{\rho\sigma}E^{\rho\sigma} +2ME_{\mu\nu}
\nonumber
\\
&+\frac{4\alpha}{3+\alpha M}\left(M_{\mu\nu}-\frac{1}{4}Mh_{\mu\nu}\right)M_{\rho\sigma}E^{\rho\sigma},
\label{App_E03}
\\
H^{(3)}_{\mu\nu}&=\frac{8}{3}\left[-N_\mu N_\nu +N^\rho\left(N_{\rho\mu\nu}+N_{\rho\nu\mu}\right) +
\frac{1}{2}N_{\rho\sigma\mu}N^{\rho\sigma}_{~~~\nu}
+N_{\mu\rho\sigma}N_{\nu}^{~\rho\sigma}\right]
\nonumber 
\\
&+
\left[2h_{\mu\nu}+\frac{8\alpha}{3(3+\alpha M)}\left(M_{\mu\nu}-\frac{1}{4}Mh_{\mu\nu}\right)\right]\left(N_\alpha N^{\alpha}-\frac{1}{2}N_{\alpha\beta\gamma}N^{\alpha\beta\gamma}\right).
\label{App_E04}
\end{align}
In all these expressions the quantities $M_{\alpha \beta \mu \nu}$, $M_{\alpha \beta}$, $M$ and $N_{\mu}$, $N_{\mu \nu \rho}$ have appeared repeatedly. These quantities have the following expressions,
\begin{align}
M_{\alpha \beta \gamma \delta}&=R_{\alpha \beta \gamma \delta}-K_{\alpha \gamma}K_{\beta \delta}+K_{\alpha \delta}K_{\beta \gamma}
\\
M_{\alpha \beta}&=R_{\alpha \beta}-KK_{\alpha \beta}K_{\alpha \gamma}K^{\gamma}_{\beta}
\\
M&=R-K^{2}+K_{\alpha \beta}K^{\alpha \beta}
\\
N_{\mu \nu \rho}&=D_{\mu}K_{\nu \rho}-D_{\nu}K_{\mu \rho}
\\
N_{\mu}&=D_{\nu}K^{\nu}_{\mu}-D_{\mu}K
\end{align}
The above expressions were used in the context of effective field equations. Let us now concentrate on the perturbative method. In this particular situation the gravitational field equations on the brane with Einstein plus \gb gravity takes the following form,
\begin{align}\label{GB_09new}
G_\mu^{\nu}&=\frac{\kappa_{5}^{2}}{\ell(1+\beta)}T_{\mu}^{~\nu}+\frac{2}{\ell}\frac{1-\beta}{1+\beta}{}^{(2)}\bar\chi_\mu^{~\nu}
+\frac{(1-3\beta)\ell^2}{1+\beta}{\cal P}_\mu^{~\nu}+\frac{\beta\ell^2}{1+\beta}C_{\mu\alpha}^{~~~\nu\beta}R_{\beta}^{~\alpha}-\frac{\beta\ell^2}{3}\left[{\cal W}_\mu^{~\nu}-\frac{7}{16}\delta_\mu^{~\nu} {\cal  W} \right].
\end{align}
Except for a few well known tensors described in the main text the other tensors appearing in the above expression turns out to be,
\begin{align}
\mathcal{W}^{\mu}_{\nu}&\equiv C_{\mu \alpha}^{~~~\beta \sigma}C_{\beta \sigma}^{~~~\nu \alpha};\qquad \mathcal{W}=C_{\mu \alpha}^{~~~\beta \sigma}C_{\beta \sigma}^{~~~\mu \alpha}
\label{Low01}
\\
\mathcal{P}_{\mu}^{\nu}&\equiv \frac{1}{6}RR^{\nu}_{\mu}-\frac{1}{4}R^{\alpha}_{\mu}R^{\nu}_{\alpha}+\frac{1}{8}\delta ^{\nu}_{\mu}R_{\alpha \beta}R^{\alpha \beta}-\frac{1}{16}\delta ^{\nu}_{\mu}R^{2}
\label{Low02}
\\
\mathcal{S}^{\nu}_{\mu}&\equiv R^{\alpha}_{\mu}R^{\nu}_{\alpha}-\frac{1}{4}\delta ^{\nu}_{\mu}R_{\alpha \beta}R^{\alpha \beta}
-\frac{1}{3}R\left(R^{\nu}_{\mu}-\frac{1}{4}\delta ^{\nu}_{\mu}R\right) -\frac{1}{2}\left(D^{\alpha}D_{\mu}R^{\nu}_{\alpha}+D_{\alpha}D^{\nu}R^{\alpha}_{\mu}\right)
\nonumber
\\
&+\frac{1}{3}D_{\mu}D^{\nu}R+\frac{1}{2}D^{2}R^{\nu}_{\mu}-\frac{1}{12}\delta ^{\nu}_{\mu}D^{2}R
\label{Low03}
\\
^{(2)}\bar{\chi}^{\nu}_{\mu}&\equiv ~^{(2)}\chi ^{\nu}_{\mu}+\frac{\ell ^{3}}{4}\mathcal{S}^{\nu}_{\mu}+\frac{\beta \ell ^{3}}{6(1-\beta)}\left(\mathcal{W}^{\nu}_{\mu}-\frac{1}{4}\delta ^{\nu}_{\mu}\mathcal{W}\right)+\frac{\ell ^{3}}{8}\Bigg[R^{\alpha}_{\mu}R^{\nu}_{\alpha}-\frac{1}{3}RR^{\nu}_{\mu}-\frac{1}{4}\delta ^{\nu}_{\mu}\left(R^{\alpha}_{\beta}R^{\beta}_{\alpha}-\frac{1}{3}R^{2}\right)\Bigg]
\label{GB_New_04}
\end{align}
Next we will discuss the field equations for spherically symmetric spacetime derived from the effective equation presented in \ref{App_E01}. 
\subsection{Effective Field Equations in General Relativity with \gb correction: Detailed Analysis}\label{GBApp_0102}

We are interested in static spherically symmetric spacetime for which the line element takes the form,
\begin{align}
ds^{2}=-e^{\nu}dt^{2}+e^{-\lambda}dr^{2}+r^{2}d\Omega ^{2}
\end{align}
For this line element one can immediately obtain the connections and hence the curvature tensor components. Using the curvature tensor components and its various contractions we obtain the following relations:
\begin{align}
\left(g^{tt}\right)^{2}\left(g^{rr}\right)^{2}\left(R_{trtr}\right)^{2}&=e^{-2\lambda} \left(\frac{\nu ''}{2}-\frac{\nu '\lambda '}{4}+\frac{\nu '^{2}}{4}\right)^{2}
\\
\left(g^{tt}\right)^{2}\left(g^{\theta \theta}\right)^{2}\left(R_{t\theta t\theta}\right)^{2}&=\left(g^{tt}\right)^{2}\left(g^{\phi \phi}\right)^{2}\left(R_{t\phi t\phi}\right)^{2}=e^{-2\lambda}\frac{\nu '^{2}}{4r^{2}}
\\
\left(g^{rr}\right)^{2}\left(g^{\theta \theta}\right)^{2}\left(R_{r\theta r\theta}\right)^{2}&=\left(g^{rr}\right)^{2}\left(g^{\phi \phi}\right)^{2}\left(R_{t\phi t\phi}\right)^{2}=e^{-2\lambda}\frac{\lambda '^{2}}{4r^{2}}
\\
\left(g^{\phi \phi}\right)^{2}\left(g^{\theta \theta}\right)^{2}\left(R_{\phi \theta \phi \theta}\right)^{2}&=\frac{\left(1-e^{-\lambda}\right)^{2}}{r^{4}}
\end{align}
For static spherically symmetric spacetime, the curvature tensor components that enter the effective field equations satisfy the following relations:
\begin{align}
R_{\alpha \beta \mu \nu}R^{\alpha \beta \mu \nu}&=4\Big\lbrace \left(g^{tt}\right)^{2}\left(g^{rr}\right)^{2}\left(R_{trtr}\right)^{2}+\left(g^{tt}\right)^{2}\left(g^{\theta \theta}\right)^{2}\left(R_{t\theta t\theta}\right)^{2}+\left(g^{tt}\right)^{2}\left(g^{\phi \phi}\right)^{2}\left(R_{t\phi t\phi}\right)^{2}
\nonumber
\\
&+\left(g^{rr}\right)^{2}\left(g^{\theta \theta}\right)^{2}\left(R_{r\theta r\theta}\right)^{2}+\left(g^{rr}\right)^{2}\left(g^{\phi \phi}\right)^{2}\left(R_{t\phi t\phi}\right)^{2}+\left(g^{\phi \phi}\right)^{2}\left(g^{\theta \theta}\right)^{2}\left(R_{\phi \theta \phi \theta}\right)^{2}\Big\rbrace
\\
R_{t\alpha \beta \mu}R^{t\alpha \beta \mu}&=2\Big\lbrace \left(g^{tt}\right)^{2}\left(g^{rr}\right)^{2}\left(R_{trtr}\right)^{2}+\left(g^{tt}\right)^{2}\left(g^{\theta \theta}\right)^{2}\left(R_{t\theta t\theta}\right)^{2}+\left(g^{tt}\right)^{2}\left(g^{\phi \phi}\right)^{2}\left(R_{t\phi t\phi}\right)^{2}\Big\rbrace
\\
R_{r\alpha \beta \mu}R^{r\alpha \beta \mu}&=2\Big\lbrace \left(g^{rr}\right)^{2}\left(g^{\theta \theta}\right)^{2}\left(R_{r\theta r\theta}\right)^{2}+\left(g^{rr}\right)^{2}\left(g^{\phi \phi}\right)^{2}\left(R_{t\phi t\phi}\right)^{2}+\left(g^{tt}\right)^{2}\left(g^{rr}\right)^{2}\left(R_{trtr}\right)^{2}\Big\rbrace
\end{align}
Thus the temporal part of the effective field equations presented in \ref{App_E01} can be simplified using the above equations and finally that leads to,
\begin{align}
G^{t}_{t}+E^{t}_{t}&=-\alpha \Big\lbrace \frac{4}{3}R_{t}\alpha \beta \mu R^{t\alpha \beta \mu}-\frac{7}{12}R_{\alpha \beta \mu \nu} R^{\alpha \beta \mu \nu}-4R^{t}_{~\mu t\nu}E^{\mu \nu}\Big\rbrace 
\nonumber
\\
&=-\alpha \Big[\frac{1}{3} \left(g^{tt}\right)^{2}\left(g^{rr}\right)^{2}\left(R_{trtr}\right)^{2} +\frac{1}{3} \left(g^{tt}\right)^{2}\left(g^{\theta \theta}\right)^{2}\left(R_{t\theta t\theta}\right)^{2} +\frac{1}{3} \left(g^{tt}\right)^{2}\left(g^{\phi \phi}\right)^{2}\left(R_{t\phi t\phi}\right)^{2} 
\nonumber
\\
&-\frac{7}{3}\left(g^{rr}\right)^{2}\left(g^{\theta \theta}\right)^{2}\left(R_{r\theta r\theta}\right)^{2} -\frac{7}{3} \left(g^{rr}\right)^{2}\left(g^{\phi \phi}\right)^{2}\left(R_{t\phi t\phi}\right)^{2} -\frac{7}{3} \left(g^{\phi \phi}\right)^{2}\left(g^{\theta \theta}\right)^{2}\left(R_{\phi \theta \phi \theta}\right)^{2}
\nonumber
\\
&-4g^{tt}R_{trtr}E^{rr}-4g^{tt}R_{t\theta t\theta}E^{\theta \theta}-4g^{tt}R_{t\phi t\phi}E^{\phi \phi}
\Big]
\label{App_E05}
\end{align}
By identical methods the radial component of the effective equations can be obtained. This also requires writing all the curvature components explicitly using earlier identities. This leads to,
\begin{align}\label{GB_13}
G^{r}_{r}+E^{r}_{r}&=-\alpha \Big[\frac{1}{3} \left(g^{tt}\right)^{2}\left(g^{rr}\right)^{2}\left(R_{trtr}\right)^{2} +\frac{1}{3} \left(g^{tt}\right)^{2}\left(g^{\theta \theta}\right)^{2}\left(R_{t\theta t\theta}\right)^{2} +\frac{1}{3} \left(g^{tt}\right)^{2}\left(g^{\phi \phi}\right)^{2}\left(R_{t\phi t\phi}\right)^{2} 
\nonumber
\\
&-\frac{7}{3}\left(g^{rr}\right)^{2}\left(g^{\theta \theta}\right)^{2}\left(R_{r\theta r\theta}\right)^{2} -\frac{7}{3} \left(g^{rr}\right)^{2}\left(g^{\phi \phi}\right)^{2}\left(R_{t\phi t\phi}\right)^{2} -\frac{7}{3} \left(g^{\phi \phi}\right)^{2}\left(g^{\theta \theta}\right)^{2}\left(R_{\phi \theta \phi \theta}\right)^{2}
\nonumber
\\
&-4g^{tt}R_{trtr}E^{rr}-4g^{tt}R_{t\theta t\theta}E^{\theta \theta}-4g^{tt}R_{t\phi t\phi}E^{\phi \phi}
\Big]
\end{align}
In the above expression the $G^{\mu}_{\nu}$ on the left hand side is for the full \EH plus \gb gravity, while the terms on the right hand side being already linear in $\alpha$ are evaluated for background \gr.

For the background \gr\ solution we have $e^{\nu}=e^{-\lambda}=1-(a/r)-(b/r^{2})$, for which we readily obtain:
\begin{align}
\nu '&=e^{\lambda}\left(\frac{a}{r^{2}}+\frac{2b}{r^{3}}\right)
\\
\nu ''&=e^{2\lambda}\left(-\frac{2a}{r^{3}}+\frac{a^{2}-6b}{r^{4}}+\frac{4ab}{r^{5}}+\frac{2b^{2}}{r^{6}}\right)
\\
\frac{1}{2}&\times e^{\nu}\left(\nu ''+\nu '^{2}\right)=-\frac{a}{r^{3}}-\frac{3b}{r^{4}}
\end{align}
which immediately leads to,
\begin{align}
\left(g^{tt}\right)^{2}\left(g^{rr}\right)^{2}\left(R_{trtr}\right)^{2}&=\left(\frac{a}{r^{3}}+\frac{3b}{r^{4}}\right)^{2}
\\
\left(g^{tt}\right)^{2}\left(g^{\theta \theta}\right)^{2}\left(R_{t\theta t\theta}\right)^{2}&=\left(g^{tt}\right)^{2}\left(g^{\phi \phi}\right)^{2}\left(R_{t\phi t\phi}\right)^{2}=\frac{1}{4}\left(\frac{a}{r^{3}}+\frac{2b}{r^{4}}\right)^{2}
\\
\left(g^{rr}\right)^{2}\left(g^{\theta \theta}\right)^{2}\left(R_{r\theta r\theta}\right)^{2}&=\left(g^{rr}\right)^{2}\left(g^{\phi \phi}\right)^{2}\left(R_{t\phi t\phi}\right)^{2}=\frac{1}{4}\left(\frac{a}{r^{3}}+\frac{2b}{r^{4}}\right)^{2}
\\
\left(g^{\phi \phi}\right)^{2}\left(g^{\theta \theta}\right)^{2}\left(R_{\phi \theta \phi \theta}\right)^{2}&=\left(\frac{a}{r^{3}}+\frac{b}{r^{4}}\right)^{2}
\end{align}
Thus the right hand side of \ref{App_E05} can be simplified and we finally obtain,
\begin{align}
-\alpha &\times \Big[\frac{1}{3} \left(g^{tt}\right)^{2}\left(g^{rr}\right)^{2}\left(R_{trtr}\right)^{2} +\frac{1}{3} \left(g^{tt}\right)^{2}\left(g^{\theta \theta}\right)^{2}\left(R_{t\theta t\theta}\right)^{2} +\frac{1}{3} \left(g^{tt}\right)^{2}\left(g^{\phi \phi}\right)^{2}\left(R_{t\phi t\phi}\right)^{2} 
\nonumber
\\
&-\frac{7}{3}\left(g^{rr}\right)^{2}\left(g^{\theta \theta}\right)^{2}\left(R_{r\theta r\theta}\right)^{2} -\frac{7}{3} \left(g^{rr}\right)^{2}\left(g^{\phi \phi}\right)^{2}\left(R_{t\phi t\phi}\right)^{2} -\frac{7}{3} \left(g^{\phi \phi}\right)^{2}\left(g^{\theta \theta}\right)^{2}\left(R_{\phi \theta \phi \theta}\right)^{2}\Big]
\nonumber
\\
&=-\alpha \Big[\frac{1}{3}\left(\frac{a}{r^{3}}+\frac{3b}{r^{4}}\right)^{2}+\frac{1}{6}\left(\frac{a}{r^{3}}+\frac{2b}{r^{4}}\right)^{2}-\frac{7}{6}\left(\frac{a}{r^{3}}+\frac{2b}{r^{4}}\right)^{2}-\frac{7}{3}\left(\frac{a}{r^{3}}+\frac{b}{r^{4}}\right)^{2}\Big]
\nonumber
\\
&=-\alpha \Big[\frac{a^{2}}{3r^{6}}+\frac{2ab}{r^{7}}+\frac{3b^{2}}{r^{8}}+\frac{a^{2}}{6r^{6}}+\frac{2ab}{3r^{7}}+\frac{2b^{2}}{3r^{8}}-\frac{7}{6}\left(\frac{a^{2}}{r^{6}}+\frac{4ab}{r^{7}}+\frac{4b^{2}}{r^{8}}\right)
\nonumber
\\
&-\frac{7}{3}\left(\frac{a^{2}}{r^{6}}+\frac{2ab}{r^{7}}+\frac{b^{2}}{r^{8}}\right)\Big]
\nonumber
\\
&=-\alpha \left(-3\frac{a^{2}}{r^{6}}-\frac{20}{3}\frac{ab}{r^{7}}-\frac{10}{3}\frac{b^{2}}{r^{8}}\right)
\end{align}
Substitution of which in \ref{App_E05} finally leads to,
\begin{align}
G^{t}_{t}&+3\bar{\kappa}U(r)-\alpha \left(3\frac{a^{2}}{r^{6}}+\frac{20}{3}\frac{ab}{r^{7}}+\frac{10}{3}\frac{b^{2}}{r^{8}}\right)-4\alpha \Big[-\bar{\kappa}e^{-\lambda}e^{-\nu}\left(\frac{a}{r^{3}}+\frac{3b}{r^{4}}\right)\left(U+2P\right)
\nonumber
\\
&+\bar{\kappa}\frac{1}{2r}\left(U-P\right)e^{\lambda -\nu}\left(\frac{a}{r^{2}}+\frac{2b}{r^{3}}\right)e^{\nu -\lambda}+\bar{\kappa}\frac{1}{2r}\left(U-P\right)e^{\lambda -\nu}\left(\frac{a}{r^{2}}+\frac{2b}{r^{3}}\right)e^{\nu -\lambda}\Big]=0
\label{App_E06}
\end{align}
Also the radial equation obtained from \ref{App_E01} can be simplified leading to,
\begin{align}
G^{r}_{r}&=-E^{r}_{r}-\alpha \Big[\frac{1}{3}\left(\frac{a}{r^{3}}+\frac{3b}{r^{4}}\right)^{2}+\frac{1}{6}\left(\frac{a}{r^{3}}+\frac{2b}{r^{4}}\right)^{2}-\frac{7}{6}\left(\frac{a}{r^{3}}+\frac{2b}{r^{4}}\right)^{2}-\frac{7}{3}\left(\frac{a}{r^{3}}+\frac{b}{r^{4}}\right)^{2}
\nonumber
\\
&+2e^{-\lambda}\left(3\bar{\kappa}U\right)\left(\frac{2a}{r^{3}}+\frac{6b}{r^{4}}\right)\left(-e^{-\nu}\right)
+4e^{-\lambda}\bar{\kappa}\left(U-P\right)\left(\frac{a}{r^{3}}+\frac{2b}{r^{4}}\right)\left(-e^{\lambda}\right)\Big]
\nonumber
\\
&=\bar{\kappa}\left(U+2P\right)+\alpha \left(3\frac{a^{2}}{r^{6}}+\frac{20}{3}\frac{ab}{r^{7}}+\frac{10}{3}\frac{b^{2}}{r^{8}}\right)+4\alpha \left[3\bar{\kappa}U\left(\frac{a}{r^{3}}+\frac{3b}{r^{4}}\right)+\bar{\kappa}\left(U-P\right)\left(\frac{a}{r^{3}}+\frac{2b}{r^{4}}\right)\right]
\label{App_E07}
\end{align}
These are the equations we have used in order to get \ref{GB_19} and \ref{GB_20} respectively.
\subsection{Effective Field Equations in f(R) gravity with \gb correction term: Detailed Analysis}
\label{GBApp_0103}

The metric for the background $f(\mathcal{R})$ gravity spacetime can be written in the following form,
\begin{align}
e^{\nu}=e^{-\lambda}=1-\frac{a}{r}-\frac{b}{r^{2}}+cr^{2}
\end{align}
Then we obtain the following identities
\begin{align}
e^{\nu}\nu '&=\frac{a}{r^{2}}+\frac{2b}{r^{3}}+2cr
\\
\nu ''&=e^{-2\nu}\left(-\frac{2a}{r^{3}}+\frac{a^{2}-6b}{r^{4}}+\frac{4ab}{r^{5}}+\frac{2b^{2}}{r^{6}}+2c-\frac{8ac}{r}-\frac{16bc}{r^{2}}-2c^{2}r^{2}\right)
\\
\frac{1}{2}&\times e^{\nu}\left(\nu ''+\nu '^{2}\right)=-\frac{a}{r^{3}}-\frac{3b}{r^{4}}+c
\end{align}
which immediately leads to,
\begin{align}
\left(g^{tt}\right)^{2}\left(g^{rr}\right)^{2}\left(R_{trtr}\right)^{2}&=\left(\frac{a}{r^{3}}+\frac{3b}{r^{4}}-c\right)^{2}
\\
\left(g^{tt}\right)^{2}\left(g^{\theta \theta}\right)^{2}\left(R_{t\theta t\theta}\right)^{2}&=\left(g^{tt}\right)^{2}\left(g^{\phi \phi}\right)^{2}\left(R_{t\phi t\phi}\right)^{2}=\frac{1}{4}\left(\frac{a}{r^{3}}+\frac{2b}{r^{4}}+2c\right)^{2}
\\
\left(g^{rr}\right)^{2}\left(g^{\theta \theta}\right)^{2}\left(R_{r\theta r\theta}\right)^{2}&=\left(g^{rr}\right)^{2}\left(g^{\phi \phi}\right)^{2}\left(R_{t\phi t\phi}\right)^{2}=\frac{1}{4}\left(\frac{a}{r^{3}}+\frac{2b}{r^{4}}+2c\right)^{2}
\\
\left(g^{\phi \phi}\right)^{2}\left(g^{\theta \theta}\right)^{2}\left(R_{\phi \theta \phi \theta}\right)^{2}&=\left(\frac{a}{r^{3}}+\frac{b}{r^{4}}-c\right)^{2}
\end{align}
Thus we obtain the following result
\begin{align}
-\alpha &\times \Big[\frac{1}{3} \left(g^{tt}\right)^{2}\left(g^{rr}\right)^{2}\left(R_{trtr}\right)^{2} +\frac{1}{3} \left(g^{tt}\right)^{2}\left(g^{\theta \theta}\right)^{2}\left(R_{t\theta t\theta}\right)^{2} +\frac{1}{3} \left(g^{tt}\right)^{2}\left(g^{\phi \phi}\right)^{2}\left(R_{t\phi t\phi}\right)^{2} 
\nonumber
\\
&-\frac{7}{3}\left(g^{rr}\right)^{2}\left(g^{\theta \theta}\right)^{2}\left(R_{r\theta r\theta}\right)^{2} -\frac{7}{3} \left(g^{rr}\right)^{2}\left(g^{\phi \phi}\right)^{2}\left(R_{t\phi t\phi}\right)^{2} -\frac{7}{3} \left(g^{\phi \phi}\right)^{2}\left(g^{\theta \theta}\right)^{2}\left(R_{\phi \theta \phi \theta}\right)^{2}\Big]
\nonumber
\\
&=-\alpha \Big[\frac{1}{3}\left(\frac{a}{r^{3}}+\frac{3b}{r^{4}}-c\right)^{2}+\frac{1}{6}\left(\frac{a}{r^{3}}+\frac{2b}{r^{4}}+2c\right)^{2}-\frac{7}{6}\left(\frac{a}{r^{3}}+\frac{2b}{r^{4}}+2c\right)^{2}-\frac{7}{3}\left(\frac{a}{r^{3}}+\frac{b}{r^{4}}-c\right)^{2}\Big]
\nonumber
\\
&=-\alpha \left(-3\frac{a^{2}}{r^{6}}-\frac{20}{3}\frac{ab}{r^{7}}-\frac{10}{3}\frac{b^{2}}{r^{8}}-\frac{20}{3}c^{2}-\frac{16}{3}\frac{bc}{r^{4}}\right)
\end{align}
Using the above identity we arrive at the temporal part of the effective field equations as,
\begin{align}
G^{t}_{t}+E^{t}_{t}&+\left[\Lambda _{4}-F(\mathcal{R})\right]h^{t}_{t}=-\alpha \left(-3\frac{a^{2}}{r^{6}}-\frac{20}{3}\frac{ab}{r^{7}}-\frac{10}{3}\frac{b^{2}}{r^{8}}-\frac{20}{3}c^{2}-\frac{16}{3}\frac{bc}{r^{4}}\right)
\nonumber
\\
&+4\alpha \Big[-\bar{\kappa}e^{-\lambda}e^{-\nu}\left(U+2P\right)\left(\frac{a}{r^{3}}+\frac{3b}{r^{4}}-c\right)+\frac{\bar{\kappa}}{2r}\left(U-P\right)e^{\lambda -\nu}\left(\frac{a}{r^{2}}+\frac{2b}{r^{3}}+2cr\right)e^{\nu -\lambda}
\nonumber
\\
&+\frac{\bar{\kappa}}{2r}\left(U-P\right)e^{\lambda -\nu}\left(\frac{a}{r^{2}}+\frac{2b}{r^{3}}+2cr\right)e^{\nu -\lambda}\Big]
\end{align}
Which after simplification leads to,
\begin{align}
G^{t}_{t}&=-3\bar{\kappa}U(r)+\alpha \left(3\frac{a^{2}}{r^{6}}+\frac{20}{3}\frac{ab}{r^{7}}+\frac{10}{3}\frac{b^{2}}{r^{8}}+\frac{20}{3}c^{2}+\frac{16}{3}\frac{bc}{r^{4}}\right)-\left[\Lambda _{4}-F(\mathcal{R})\right]
\nonumber
\\
&+4\alpha \Big[-\frac{3\bar{\kappa}P_{0}}{2r^{4}}\left(\frac{a}{r^{3}}+\frac{3b}{r^{4}}-c\right)
-\frac{3\bar{\kappa}P_{0}}{2r^{4}}\left(\frac{a}{r^{3}}+\frac{2b}{r^{4}}+2c\right)\Big]
\nonumber
\\
&=-3\bar{\kappa}U(r)+\alpha \left(3\frac{a^{2}}{r^{6}}+\frac{20}{3}\frac{ab}{r^{7}}+\frac{10}{3}\frac{b^{2}}{r^{8}}+\frac{20}{3}c^{2}+\frac{16}{3}\frac{bc}{r^{4}}\right)
\nonumber
\\
&-\left[\Lambda _{4}-F(\mathcal{R})\right]+\frac{6\alpha\bar{\kappa}P_{0}}{r^{4}}\left(\frac{2a}{r^{3}}+\frac{5b}{r^{4}}+c\right)
\label{App_E08}
\end{align}
which leads to the following solution for $e^{-\lambda}$ as,
\begin{align}
e^{-\lambda}&=1-\frac{2GM+Q_{0}}{r}-\frac{3\bar{\kappa}P_{0}}{2r^{2}}+\alpha \left(-\frac{a^{2}}{r^{4}}-\frac{5}{3}\frac{ab}{r^{5}}-\frac{2}{3}\frac{b^{2}}{r^{6}}+20c^{2}r^{2}-\frac{16}{3}\frac{bc}{r^{2}}\right)
\nonumber
\\
&+\left[F(\mathcal{R})-\Lambda _{4}\right]\frac{r^{2}}{3}+6\alpha \bar{\kappa}P_{0}\left(\frac{a}{2r^{5}}+\frac{b}{r^{6}}+\frac{c}{r^{2}}\right)
\label{App_E09}
\end{align}
The radial part of effective equation turns out to be,
\begin{align}
G^{r}_{r}&+E^{r}_{r}+\Lambda _{4}-F(\mathcal{R})=-\alpha \Big[\frac{1}{3}\left(\frac{a}{r^{3}}+\frac{3b}{r^{4}}-c\right)^{2}-\left(\frac{a}{r^{3}}+\frac{2b}{r^{4}}+2c\right)^{2}-\frac{7}{3}\left(\frac{a}{r^{3}}+\frac{b}{r^{4}}-c\right)^{2}\Big]
\nonumber
\\
&-\alpha \Big[-6e^{-\lambda}e^{-\nu}\bar{\kappa}U\left(\frac{2a}{r^{3}}+\frac{6b}{r^{4}}-2c\right)-4\bar{\kappa}\left(U-P\right) \left(\frac{a}{r^{3}}+\frac{2b}{r^{4}}+2c\right)\Big]
\end{align}
which ultimately leads to,
\begin{align}
G^{r}_{r}&=\left[F(\mathcal{R})-\Lambda _{4}\right]+\bar{\kappa}\left(U+2P\right)+\alpha \Big[\frac{3a^{2}}{r^{6}}+\frac{10}{3}\frac{b^{2}}{r^{8}}+\frac{20}{3}c^{2}+\frac{20}{3}\frac{ab}{r^{7}}+\frac{10}{3}\frac{bc}{r^{4}}\Big]
\nonumber
\\
&+4\alpha \left[3\bar{\kappa}U(r)\left(\frac{a}{r^{3}}+\frac{3b}{r^{4}}-c\right)+\bar{\kappa}\left(U-P\right)\left(\frac{a}{r^{3}}+\frac{2b}{r^{4}}+2c\right)\right]
\label{App_E10}
\end{align}
This again shows that $e^{\nu}=e^{-\lambda}$.
\subsection{Low Energy Effective Action for General Relativity with \gb correction term: Detailed Analysis}
\label{GBApp_0104}

The temporal component of the low energy effective field equations, as presented in \ref{GB_09} reads,
\begin{align}
G^{t}_{t}&=\frac{2}{\ell}\frac{1-\beta}{1+\beta}~ ^{(2)}\chi ^{t}_{t}+\frac{2\beta \ell ^{2}}{1+\beta}\left(R_{t\alpha \beta \mu}R^{t\alpha \beta \mu}-\frac{1}{4}R_{\alpha \beta \mu \nu}R^{\alpha \beta \mu \nu}\right)
\nonumber
\\
&-\frac{\beta \ell ^{2}}{3}\left(R_{t\alpha \beta \mu}R^{t\alpha \beta \mu}-\frac{7}{16}R_{\alpha \beta \mu \nu}R^{\alpha \beta \mu \nu}\right)
\nonumber
\\
&=\frac{2}{\ell}\frac{1-\beta}{1+\beta}~ ^{(2)}\chi ^{t}_{t}+\frac{(23-\beta)\beta \ell ^{2}}{12(1+\beta)}\left(R_{trtr}R^{trtr}+R_{t\theta t\theta}R^{t\theta t\theta}+R_{t\phi t\phi}R^{t\phi t\phi}\right)
\nonumber
\\
&+\frac{(7\beta -17)\beta \ell ^{2}}{12(1+\beta)}\left(R_{r\theta r\theta}R^{r\theta r\theta}+R_{r\phi r\phi}R^{r\phi r\phi}+R_{\theta \phi \theta \phi}R^{\theta \phi \theta \phi}\right)
\end{align}
Again starting from the expressions for curvature components with metric elements $e^{\nu}=e^{-\lambda}=1-(a/r)-(b/r^{2})$, we readily obtain,
\begin{align}
G^{t}_{t}&=\frac{2}{\ell}\frac{1-\beta}{1+\beta}~ ^{(2)}\chi ^{t}_{t}+\frac{(23-\beta)\beta \ell ^{2}}{12(1+\beta)}\left[\left(\frac{a}{r^{3}}+\frac{3b}{r^{4}}\right)^{2}+\frac{1}{2}\left(\frac{a}{r^{3}}+\frac{2b}{r^{4}}\right)^{2}\right]
\nonumber
\\
&+\frac{(7\beta -17)\beta \ell ^{2}}{12(1+\beta)}\left[\frac{1}{2}\left(\frac{a}{r^{3}}+\frac{2b}{r^{4}}\right)^{2}+\left(\frac{a}{r^{3}}+\frac{b}{r^{4}}\right)^{2}\right]
\end{align}
Substituting for the expression for $G^{t}_{t}$, we get,
\begin{align}
\frac{1}{r^{2}}\dfrac{d}{dr}\left(re^{-\lambda}\right)&-\frac{1}{r^{2}}=\frac{2}{\ell}\frac{1-\beta}{1+\beta}~ ^{(2)}\chi ^{t}_{t}+\frac{(23-\beta)\beta \ell ^{2}}{12(1+\beta)}\left(\frac{3}{2}\frac{a^{2}}{r^{6}}+\frac{8ab}{r^{7}}+\frac{11b^{2}}{r^{8}}\right)
\nonumber
\\
&+\frac{(7\beta -17)\beta \ell ^{2}}{12(1+\beta)}\left(\frac{3a^{2}}{2r^{6}}+\frac{4ab}{r^{7}}+\frac{3b^{2}}{r^{8}}\right)
\end{align}
which can be integrated to yield,
\begin{align}
e^{-\lambda}&=1-\frac{2GM+Q_{0}}{r}+\frac{2}{\ell}\frac{1-\beta}{1+\beta}\frac{1}{r}~\int ~^{(2)}\chi ^{t}_{t}(r)r^{2}dr
+\frac{(23-\beta)\beta \ell ^{2}}{12(1+\beta)}\left(-\frac{a^{2}}{2r^{4}}-\frac{2ab}{r^{5}}-\frac{11b^{2}}{5r^{6}}\right)
\nonumber
\\
&+\frac{(7\beta -17)\beta \ell ^{2}}{12(1+\beta)}\left(-\frac{a^{2}}{2r^{4}}-\frac{ab}{r^{5}}-\frac{3b^{2}}{5r^{6}}\right)
\end{align}
The radial equation leads to,
\begin{align}
G^{r}_{r}&=\frac{2}{\ell}\frac{1-\beta}{1+\beta}~ ^{(2)}\chi ^{r}_{r}+\frac{(23-\beta)\beta \ell ^{2}}{12(1+\beta)}\left(R_{r\theta r\theta}R^{r\theta r\theta}+R_{rtrt}R^{rtrt}+R_{r\phi r\phi}R^{r\phi r\phi}\right)
\nonumber
\\
&+\frac{(7\beta -17)\beta \ell ^{2}}{12(1+\beta)}\left(R_{t\theta t\theta}R^{t\theta t\theta}+R_{t\phi t\phi}R^{t\phi t\phi}+R_{\theta \phi \theta \phi}R^{\theta \phi \theta \phi}\right)
\nonumber
\\
&=\frac{2}{\ell}\frac{1-\beta}{1+\beta}~ ^{(2)}\chi ^{r}_{r}+\frac{(23-\beta)\beta \ell ^{2}}{12(1+\beta)}\left(\frac{3}{2}\frac{a^{2}}{r^{6}}+\frac{8ab}{r^{7}}+\frac{11b^{2}}{r^{8}}\right)
\nonumber
\\
&+\frac{(7\beta -17)\beta \ell ^{2}}{12(1+\beta)}\left(\frac{3a^{2}}{2r^{6}}+\frac{4ab}{r^{7}}+\frac{3b^{2}}{r^{8}}\right)
\end{align}
hence $e^{\nu}=e^{-\lambda}$. However from the above structure the corrections are not immediately obvious. For that we note using $^{(2)}E_{\mu \nu}=-(2/\ell)~^{(2)}\bar{\chi}_{\mu \nu}$ the field equations transform to, 
\begin{align}
G^{\rho}_{\sigma}&=\frac{2}{\ell}\frac{1-\beta}{1+\beta}~ ^{(2)}\bar{\chi}^{\rho}_{\sigma}-\frac{\beta \ell ^{2}}{3}\left(R_{\sigma \alpha \beta \mu}R^{\rho \alpha \beta \mu}-\frac{7}{16}\delta ^{\rho}_{\sigma} R_{\alpha \beta \mu \nu}R^{\alpha \beta \mu \nu}\right)
\nonumber
\\
&=-E^{\rho}_{\sigma}+\frac{4\beta}{\ell (1+\beta)}~^{(2)}\chi ^{\rho}_{\sigma}+\frac{4\beta ^{2}\ell ^{2}}{6(1-\beta ^{2})}\left(R_{\sigma \alpha \beta \mu}R^{\rho \alpha \beta \mu}-\frac{1}{4}\delta ^{\rho}_{\sigma} R_{\alpha \beta \mu \nu}R^{\alpha \beta \mu \nu}\right)
\nonumber
\\
&-\frac{\beta \ell ^{2}}{3}\left(R_{\sigma \alpha \beta \mu}R^{\rho \alpha \beta \mu}-\frac{7}{16}\delta ^{\rho}_{\sigma} R_{\alpha \beta \mu \nu}R^{\alpha \beta \mu \nu}\right)
\end{align}
Thus keeping terms upto linear order in $\beta$ the temporal component turns out to be,
\begin{align}\label{App_E11}
G^{t}_{t}+E^{t}_{t}&=\frac{4\beta}{\ell (1+\beta)}~^{(2)}\chi ^{t}_{t}-\frac{\beta \ell ^{2}}{12}\left(R_{trtr}R^{trtr}+R_{t\theta t\theta}R^{t\theta t\theta}+R_{t\phi t\phi}R^{t\phi t\phi}\right)
\nonumber
\\
&+\frac{7\beta \ell ^{2}}{12}\left(R_{r\theta r\theta}R^{r\theta r\theta}+R_{r\phi r\phi}R^{r\phi r\phi}+R_{\theta \phi \theta \phi}R^{\theta \phi \theta \phi}\right)
\nonumber
\\
&=\frac{4\beta}{\ell (1+\beta)}~^{(2)}\chi ^{t}_{t}-\frac{\beta \ell ^{2}}{12}\left(\frac{3}{2}\frac{a^{2}}{r^{6}}+\frac{8ab}{r^{7}}+\frac{11b^{2}}{r^{8}}\right)
+\frac{7\beta \ell ^{2}}{12}\left(\frac{3a^{2}}{2r^{6}}+\frac{4ab}{r^{7}}+\frac{3b^{2}}{r^{8}}\right)
\end{align}
From which the metric element $e^{-\lambda}$ would turn out to be,
\begin{align}
e^{-\lambda}&=1-\frac{2GM+Q_{0}}{r}-\frac{3\bar{\kappa}P_{0}}{2r^{2}}+\frac{4\beta}{\ell (1+\beta)}\frac{1}{r}\int ~^{(2)}\chi ^{t}_{t}~r^{2}dr-\frac{\beta \ell ^{2}}{12}\left(3\frac{a^{2}}{r^{4}}+\frac{5ab}{r^{5}}+\frac{2b^{2}}{r^{6}}\right)
\label{App_E12}
\end{align}
We will turn to the last bit of derivation, which involves the low energy equation with $f(\mathcal{R})$ gravity in the bulk.
\subsection{Low Energy Effective Action for f(R) gravity with \gb correction term: Detailed Analysis}
\label{GBApp_0105}

The low energy effective equation with $f(\mathcal{R})$ gravity corresponds to,
\begin{align}
G^{\rho}_{\sigma}&=\left[F(\mathcal{R})-\Lambda _{4}\right]\delta ^{\rho}_{\sigma}+\frac{2}{\ell}\frac{1-\beta}{1+\beta}~ ^{(2)}\chi ^{\rho}_{\sigma}+\frac{2\beta \ell ^{2}}{1+\beta}\left(R_{\sigma \alpha \beta \mu}R^{\rho \alpha \beta \mu}-\frac{1}{4}R_{\alpha \beta \mu \nu}R^{\alpha \beta \mu \nu}\right)
\nonumber
\\
&-\frac{\beta \ell ^{2}}{3}\left(R_{\sigma \alpha \beta \mu}R^{\rho \alpha \beta \mu}-\frac{7}{16}R_{\alpha \beta \mu \nu}R^{\alpha \beta \mu \nu}\right)
\end{align}
Now considering the temporal component for background metric element $e^{\nu}=e^{-\lambda}=1-(a/r)-(b/r^{2})+cr^{2}$ we arrive at,
\begin{align}
G^{t}_{t}&=\left[F(\mathcal{R})-\Lambda _{4}\right]+\frac{2}{\ell}\frac{1-\beta}{1+\beta}~ ^{(2)}\chi ^{t}_{t}+\frac{(23-\beta)\beta \ell ^{2}}{12(1+\beta)}\left(R_{trtr}R^{trtr}+R_{t\theta t\theta}R^{t\theta t\theta}+R_{t\phi t\phi}R^{t\phi t\phi}\right)
\nonumber
\\
&+\frac{(7\beta -17)\beta \ell ^{2}}{12(1+\beta)}\left(R_{r\theta r\theta}R^{r\theta r\theta}+R_{r\phi r\phi}R^{r\phi r\phi}+R_{\theta \phi \theta \phi}R^{\theta \phi \theta \phi}\right)
\nonumber
\\
&=\left[F(\mathcal{R})-\Lambda _{4}\right]+\frac{2}{\ell}\frac{1-\beta}{1+\beta}~ ^{(2)}\chi ^{t}_{t}+\frac{(23-\beta)\beta \ell ^{2}}{12(1+\beta)}\left[\left(\frac{a}{r^{3}}+\frac{3b}{r^{4}}-c\right)^{2}+\frac{1}{2}\left(\frac{a}{r^{3}}+\frac{2b}{r^{4}}+2c\right)^{2}\right]
\nonumber
\\
&+\frac{(7\beta -17)\beta \ell ^{2}}{12(1+\beta)}\left[\frac{1}{2}\left(\frac{a}{r^{3}}+\frac{2b}{r^{4}}+2c\right)^{2}+\left(\frac{a}{r^{3}}+\frac{b}{r^{4}}-c\right)^{2}\right]
\end{align}
writing explicitly we obtain,
\begin{align}
\frac{1}{r^{2}}\dfrac{d}{dr}\left(re^{-\lambda}\right)-\frac{1}{r^{2}}&=\left[F(\mathcal{R})-\Lambda _{4}\right]+\frac{2}{\ell}\frac{1-\beta}{1+\beta}~ ^{(2)}\chi ^{t}_{t}+\frac{(23-\beta)\beta \ell ^{2}}{12(1+\beta)}\Big[\frac{3a^{2}}{2r^{6}}+\frac{11b^{2}}{r^{8}}+\frac{8ab}{r^{7}}+3c^{2}-\frac{2bc}{r^{4}}\Big]
\nonumber
\\
&+\frac{(7\beta -17)\beta \ell ^{2}}{12(1+\beta)}\Big[\frac{3a^{2}}{2r^{6}}+\frac{3b^{2}}{r^{8}}+\frac{4ab}{r^{7}}+3c^{2}-\frac{2bc}{r^{4}}\Big]
\end{align}
The metric element can be solved as,
\begin{align}
e^{-\lambda}&=1-\frac{2GM+Q_{0}}{r}+\frac{F(\mathcal{R})-\Lambda _{4}}{3}r^{2}+\frac{2}{\ell}\frac{1-\beta}{1+\beta} \frac{1}{r}\int ~ ^{(2)}\chi ^{t}_{t}~r^{2}dr
\nonumber
\\
&+\frac{(23-\beta)\beta \ell ^{2}}{12(1+\beta)}\Big[-\frac{a^{2}}{2r^{4}}-\frac{11b^{2}}{5r^{6}}-\frac{2ab}{r^{5}}+c^{2}r^{2}-\frac{2bc}{r^{2}}\Big]
\nonumber
\\
&+\frac{(7\beta -17)\beta \ell ^{2}}{12(1+\beta)}\Big[-\frac{a^{2}}{2r^{4}}-\frac{3b^{2}}{5r^{6}}-\frac{ab}{r^{5}}+c^{2}r^{2}-\frac{2bc}{r^{2}}\Big]
\end{align}
In this case as well we note, $^{(2)}E^{\mu}_{\nu}=-(2/\ell)~^{(2)}\bar{\chi}^{\mu}_{\nu}$. Then,
\begin{align}
G^{\rho}_{\sigma}&=\left[F(\mathcal{R})-\Lambda _{4}\right]\delta ^{\rho}_{\sigma}+\frac{2}{\ell}\frac{1-\beta}{1+\beta}~ ^{(2)}\bar{\chi}^{\rho}_{\sigma}-\frac{\beta \ell ^{2}}{3}\left(R_{\sigma \alpha \beta \mu}R^{\rho \alpha \beta \mu}-\frac{7}{16}\delta ^{\rho}_{\sigma} R_{\alpha \beta \mu \nu}R^{\alpha \beta \mu \nu}\right)
\nonumber
\\
&=-E^{\rho}_{\sigma}+\left[F(\mathcal{R})-\Lambda _{4}\right]\delta ^{\rho}_{\sigma}+\frac{4\beta}{\ell (1+\beta)}~^{(2)}\chi ^{\rho}_{\sigma}+\frac{4\beta ^{2}\ell ^{2}}{6(1-\beta ^{2})}\left(R_{\sigma \alpha \beta \mu}R^{\rho \alpha \beta \mu}-\frac{1}{4}\delta ^{\rho}_{\sigma} R_{\alpha \beta \mu \nu}R^{\alpha \beta \mu \nu}\right)
\nonumber
\\
&-\frac{\beta \ell ^{2}}{3}\left(R_{\sigma \alpha \beta \mu}R^{\rho \alpha \beta \mu}-\frac{7}{16}\delta ^{\rho}_{\sigma} R_{\alpha \beta \mu \nu}R^{\alpha \beta \mu \nu}\right)
\end{align}
Thus keeping terms upto linear order in $\beta$ the temporal component turns out to be,
\begin{align}\label{App_E13}
G^{t}_{t}+E^{t}_{t}&=\left[F(\mathcal{R})-\Lambda _{4}\right]+\frac{4\beta}{\ell (1+\beta)}~^{(2)}\chi ^{t}_{t}-\frac{\beta \ell ^{2}}{12}\left(R_{trtr}R^{trtr}+R_{t\theta t\theta}R^{t\theta t\theta}+R_{t\phi t\phi}R^{t\phi t\phi}\right)
\nonumber
\\
&+\frac{7\beta \ell ^{2}}{12}\left(R_{r\theta r\theta}R^{r\theta r\theta}+R_{r\phi r\phi}R^{r\phi r\phi}+R_{\theta \phi \theta \phi}R^{\theta \phi \theta \phi}\right)
\nonumber
\\
&=\left[F(\mathcal{R})-\Lambda _{4}\right]+\frac{4\beta}{\ell (1+\beta)}~^{(2)}\chi ^{t}_{t}-\frac{\beta \ell ^{2}}{12}\left[\left(\frac{a}{r^{3}}+\frac{3b}{r^{4}}-c\right)^{2}+\frac{1}{2}\left(\frac{a}{r^{3}}+\frac{2b}{r^{4}}+2c\right)^{2}\right]
\nonumber
\\
&+\frac{7\beta \ell ^{2}}{12}\left[\frac{1}{2}\left(\frac{a}{r^{3}}+\frac{2b}{r^{4}}+2c\right)^{2}+\left(\frac{a}{r^{3}}+\frac{b}{r^{4}}-c\right)^{2}\right]
\nonumber
\\
&=\left[F(\mathcal{R})-\Lambda _{4}\right]+\frac{4\beta}{\ell (1+\beta)}~^{(2)}\chi ^{t}_{t}-\frac{\beta \ell ^{2}}{12}\Big[\frac{3a^{2}}{2r^{6}}+\frac{11b^{2}}{r^{8}}+\frac{8ab}{r^{7}}+3c^{2}-\frac{2bc}{r^{4}}\Big]
\nonumber
\\
&+\frac{7\beta \ell ^{2}}{12}\Big[\frac{3a^{2}}{2r^{6}}+\frac{3b^{2}}{r^{8}}+\frac{4ab}{r^{7}}+3c^{2}-\frac{2bc}{r^{4}}\Big]
\end{align}
For which the metric element turns out to be,
\begin{align}\label{App_E14}
e^{-\lambda}&=1-\frac{2GM+Q_{0}}{r}-\frac{3\bar{\kappa}P_{0}}{2r^{2}}+\frac{F(\mathcal{R})-\Lambda _{4}}{3}r^{2}+\frac{4\beta}{\ell (1+\beta)}\frac{1}{r}\int ~^{(2)}\chi ^{t}_{t}~r^{2}dr
\nonumber
\\
&-\frac{\beta \ell ^{2}}{12}\Big[\frac{3a^{2}}{r^{4}}+\frac{2b^{2}}{r^{6}}+\frac{5ab}{r^{5}}-6c^{2}r^{2}-\frac{12bc}{r^{2}}\Big]
\nonumber
\\
&=1-\frac{2GM+Q_{0}}{r}-\frac{3\bar{\kappa}P_{0}}{2r^{2}}+\frac{F(\mathcal{R})-\Lambda _{4}}{3}r^{2}+\frac{16 \alpha}{\ell ^{3}}\frac{1}{r}\int ~^{(2)}\chi ^{t}_{t}~r^{2}dr
\nonumber
\\
&-\frac{\alpha}{3}\Big[\frac{3a^{2}}{r^{4}}+\frac{2b^{2}}{r^{6}}+\frac{5ab}{r^{5}}-6c^{2}r^{2}-\frac{12bc}{r^{2}}\Big]
\end{align}
\bibliography{Brane,Gravity_1_full,Gravity_2_partial}

\bibliographystyle{./utphys1}
\end{document}